\documentclass[preprint2]{aastex}
\usepackage{natbib}

\usepackage{graphicx,graphics,subfigure}

\bibpunct{(}{)}{;}{a}{}{,}

\slugcomment{To appear in the Astronomical Journal}

\shorttitle{{\em Spitzer} observations of Coma}
\shortauthors{Edwards et al.}

\begin{document}

\title{A multi-wavelength analysis of {\it Spitzer} selected Coma Cluster galaxies: star formation rates and masses}

\author{Louise O. V. Edwards$^{1,2}$ and Dario Fadda$^{2}$
\affil{$^{1}$Department of Physics, Mount Allison University, Sackville, NB, Canada E4L 1E6}}
\affil{$^{2}$NASA Herschel Science Center, Caltech 100-22, Pasadena, CA 91125}

\email{ledwards@mta.ca}
\email{fadda@ipac.caltech.edu}

\begin{abstract}

We present a thorough study of the specific star formation rates for MIPS 24$\mu$m selected galaxies in the Coma cluster. We build galaxy spectral energy distributions using optical (u$^{\prime}$,g$^{\prime}$,r$^{\prime}$,i$^{\prime}$,z$^{\prime}$), Near-infrared (J,H,K$_{s}$), and Mid to Far-infrared (IRAC and MIPS) photometry. New and archival spectra confirm 210 cluster members. Subsequently, the total infrared luminosity, galaxy stellar mass, and specific star formation rate for the members are determined by measuring best-fit templates.  Using an array of complementary diagnostics, we search for contaminating AGN, but find few. We compare obscured star formation rates to unobscured rates derived from extinction-corrected H$\alpha$ emission line measurements. The agreement between these two values leads us to conclude that there is no evidence for an additionally obscured component. In our spectroscopic sample, complete to 80\% for r$^{\prime}$$<$ 19.5, we find that all starbursts are blue and are dwarfs, having masses $<$10$^{9}$M$_{\odot}$. Examining the location of these starbursts within the cluster, we confirm that there is a lower fraction in the cluster core. 

\end{abstract}

\keywords{galaxies: clusters: individual (Coma Cluster)  -- infrared: galaxies -- galaxies:fundamental parameters (luminosities, masses)}

\section{Introduction}

Galaxy clusters are dynamic systems in which tens to hundreds of galaxies are plowing through hot intracluster gas. In the core of the cluster,  the galaxy density is the highest, and star forming galaxies preferentially avoid these regions \citep{ken83,dre99,gav09}. However, at the outskirts of the cluster many galaxies show starburst activity detected at Mid-Infrared (IR) \citep{koy10,gal09} and radio \citep{hec85} wavelengths, as well as in optical emission lines \citep{rin05}.

Coma, at z$\sim$0.023, is our nearest rich galaxy cluster. As such, it allows for a level of detailed study currently unavailable for more distant clusters. \citet{mob03}  studied the R-band luminosity function, finding no difference in the evolutionary history of galaxies at the outskirts compared to those at the cluster core. However, the study was based on a spectroscopic sample and less deep than other studies. More recent studies, however, have revealed a segregation of galaxy activity with local density: Galaxies at the cluster center are oldest \citep{pog01,smi08}, with the fraction of blue, spiral, \citep{ter01,agu04,biv02},  and dwarf \citep{ode02,gav10} galaxies increasing with distance from the core. In the radio, a luminosity function has been well studied down to a very low surface brightnesses \citep{mil09}. These faint radio galaxies are vast in number, have high specific star formation rates (sSFRs), strong optical lines, and appear to have undergone a recent burst of star formation, prior to their quenching.  Metallicity gradients in cluster galaxies have also been noted, and linked to the age segregation, as well as to pressure confinement from the deep potential at the cluster core \citep{pog01,car02}. 

A scenario in which the galaxy activity is quenched during infall toward the dense cluster core is supported by galaxy alignments \citep{ada09}, alpha element gradients \citep{smi09}, sizes of spiral disks \citep{agu04}, and the correlation of post starburst galaxy positions with the location of X-ray substructures \citep{pog04} and infall regions \citep{mah11}. A population of recently quenched elliptical galaxies would also explain the observations of  \citet{cle09} who studied the color-magnitude diagram  of Coma at Mid-IR wavelengths finding that 68\% of ellipticals were truly passive, but 32\% showed evidence for very recent activity. 

This picture is consistent with results from  cosmological simulations that show the build-up of massive clusters through accretion of infalling galaxies and galaxy groups. The numerical simulations of \citet{bek99} show higher star formation rates (SFRs) in the accreted galaxies, in agreement with recent observational results. \citet{gal09} found that most of the obscured star formation in the complex Abell~901-Abell~902 comes from galaxies between and at the edges of the two clusters.  Our work on Abell~1763 \citep{edw10,fad08} revealed a $\sim$12~Mpc filament of Mid-IR emitting galaxies connecting this cluster to the nearby Abell~1770. For that case, the fraction of star forming filament galaxies was found to be double that in the cluster regions.

Work by \citet{jen07},   \citet{bai06} and \citet{mah10}  has shown that the dwarf star forming galaxy population in Coma is quenched in the high density core. However, the dwarf galaxies do not necessarily have very intense starbursts, therefore, a study of the {\it specific} star formation rates is important. This requires an accurate method for determining galaxy stellar mass determinations. Thus, of particular importance to this study are our new Near-IR observations as they bracket the stellar mass bump, allowing for accurate stellar mass measurements. In this paper, we combine these original Near-IR photometry with public Mid-Infrared photometry from the IRAC and MIPS instruments aboard the  {\it Spitzer} Space Telescope, in addition to optical photometry from the Sloan Digital Sky Survey (SDSS). We gather optical spectra from the SDSS as well as from our own campaign. With these data we build spectral energy distributions (SEDs) of Coma cluster members. Star formation rates are calculated from the total infrared luminosity derived from best-fit template galaxy SEDs. 

In the next section, we present the observations and reduction methods for the deep Near-IR J, H, and K$_{s}$ data which was collected from the Palomar WIRC instrument, as well as the {\it Spitzer} MIPS 24, 70, and 160$\mu$m and IRAC 3.6, 4.5, 5.8, and 8.0$\mu$m data which we have re-reduced. We also include the optical photometry from the SDSS Data Release~7, the available spectra from that program, as well as 338 spectra of 24$\mu$m selected galaxies we gathered from the HYDRA instrument on the WIYN telescope. In Section 3, we present the catalog of 24$\mu$m selected galaxies in Coma, along with the associated optical and infrared fluxes and redshifts. We release our fully reduced images and spectra to the public via the NASA/IPAC Infrared Science Archive (IRSA) and the NASA Extragalactic Database (NED). These data allow us to determine the amount of AGN contamination, measure stellar mass, and derive the total infrared luminosities  required to calculate specific star formation rates (SFRs), which we present Section 4. We compare these obscured SFRs to unobscured rates measured from nebular emission lines in Section 5, where we also compare other proxies for the obscured SFR and discuss the specific SFR as a function of location. We conclude, in Section 6, that the starburst galaxies are blue dwarfs and that there is a paucity of these in the dense cluster core. This is in line with the hypothesis that starburst galaxies have been recently accreted onto the cluster and will be quenched as they make their way to the core.

\section{Observations and Data Reduction}

\begin{deluxetable}{lccccccc}
\tabletypesize{\scriptsize}
\tablewidth{0pt}
%\tablecolumns{<num>}
\tablecaption{Observations \label{obs}}
\tablehead{\colhead{INST} & \colhead{$\lambda$$_{cent}$ ($\mu$m)} & \colhead{ID} & \colhead{Date} & \colhead{Time(min)}& \colhead{Coverage($^{\prime 2}$)} & \colhead{Depth} &\colhead{FWHM} ($^{\prime\prime}$)}
\startdata
IRAC & 3.6, 4.5, 5.8, 8.0 &r3859712 &   2004 Jul 04 & 160.1 & 3250 & 1.5, 4.0, 3.9, 3.8 $\mu$Jy 5$\sigma$& 1.7,1.7,1.9,2.0 \\
IRAC & 3.6, 4.5, 5.8, 8.0 &r3859456 &   2003 dec 18 & 110.9 &  && \\
IRAC & 3.6, 4.5, 5.8, 8.0 &r3859968 &   2004 Jul 04 & 160.1 &  && \\
MIPS & 24, 70, 160 &r4740096&   2004 Jun 22 & 169.6 &11070, 10530& 0.08, 32.0, 130 mJy 5$\sigma$& 5.9, 16, 36\\
MIPS & 24, 70, 160 &r4740352&   2004 Jun 22 & 169.6 &  &&\\
MIPS & 24, 70, 160 &r18315776&   2007 Jan 19 & 170.3 &  &&\\
MIPS & 24, 70, 160 &r18316032&   2007 Jan 19 & 170.3 &  &&\\
MIPS & 24, 70, 160 &r18315520&   2007 Jan 19 & 170.3 &  &&\\
MIPS & 24, 70, 160 &r18316288&   2007 Jan 19 & 170.3 &  &&\\
\hline
\hline
WIRCJ & 1.250 &     ...     &  2008 Apr 23 & 250.0& 8480 & 19.5 mag&1.2\\
WIRCH & 1.635  &    ...     &  2008 Apr 25 & 187.0&8480   & 18.5 mag&0.9\\
WIRCK$_{s}$ & 2.150   &    ...    &  2008 Apr 24 & 247.0& 8480 & 17.0 mag&1.3\\
\enddata
\end{deluxetable}

\begin{figure*}
\epsscale{1.8}
\plotone{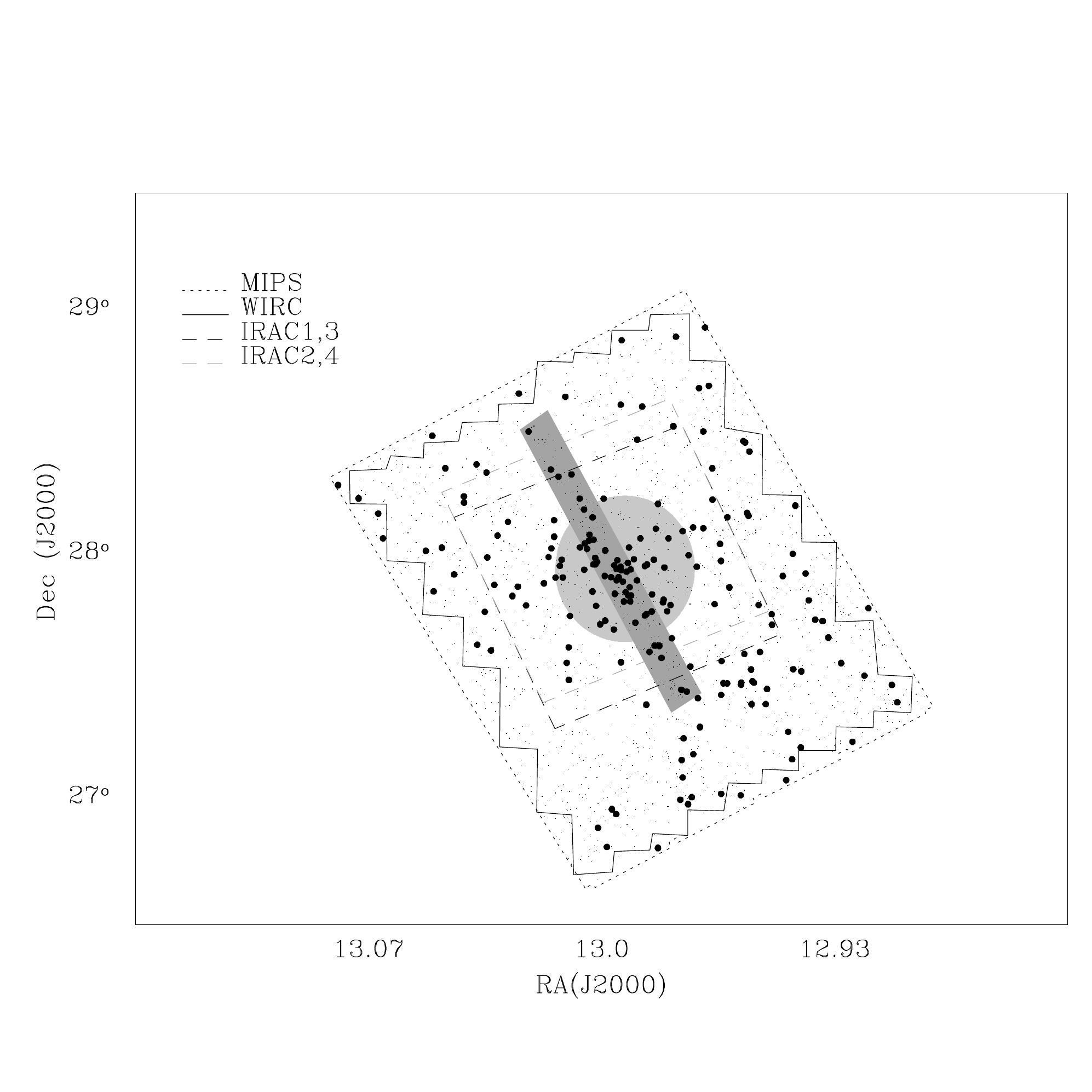}
\caption{{\bf Coverage of IR and optical data.}  The coverage of the infrared photometry. MIPS 24$\mu$m coverage is represented by the large dotted rectangle, IRAC 3.6 and 5.8$\mu$m fields are shown as smaller dashed squares in black and grey, respectively. The black polygon shows the Near-IR coverage. SDSS galaxies with r$^{\prime}<$ 20.5 and MIPS emission are shown as points, and those that are spectroscopically confirmed cluster members are large black dots. The filled grey circle has a radius of 0.3$\,$degrees and highlights the cluster core (0.2$\times$r$_{200}$). The MIPS data cover 8480~arcmin$^{2}$. Deeper 24$\mu$m observations exist over the area covered by the grey rectangle. \label{spitzer_cov}
}
\end{figure*}

In this section, we describe the observations and data reduction. Table~\ref{obs} lists our observational dataset and Figure~\ref{spitzer_cov} shows the coverage of the infrared data and the location of spectroscopic members.

The photometric data was acquired from recent WIRC Near-IR and archival Mid and Far-IR {\it Spitzer} imaging (PIs: Fazio, Rieke, Kitayama), optical catalogs from the SDSS DR7, and radio source catalogs from the Faint Images of the Radio Sky at Twenty-Centimeters (FIRST). Because SED fitting is more accurate when the MIPS 24$\mu$m datapoint exists, we include only galaxies which are detected at 24$\mu$m to at least the 3$\sigma$ level. The MIPS data field of view covers the full core of the cluster, (0.2 - 0.5)$\times$r$_{200}$ (where r$_{200}$ is the radius within which the cluster density is 200 times the mean density required to close the universe). This same region has been mapped in J, H, and K$_{s}$ using the WIRC instrument on the Palomar 200$\,$in telescope. 

We also take advantage of spectra from SDSS DR7 which we add to our own data from WIYN.

\subsection{WIRC Near-IR Observations}

The Near-IR magnitudes are driven by old stars, which comprise the bulk of the galaxy mass and are therefore paramount in our later derivation of the galaxy stellar mass.

The Near-IR photometry was gathered at the Palomar 200in telescope using the Widefield InfraRed Camera (WIRC) during three nights of observations between April 23-25, 2008. As detailed in Table~1, the nights were clear with seeing between 0.9-1.3$^{\prime\prime}$ and most of the region covered by {\it Spitzer} was imaged, as shown in Figure~\ref{spitzer_cov}. Fields in J were observed in sequences of 40s, and for 30s in H. In K$_{s}$, 3 coadds of 13s each were taken to avoid saturation. The total integration times for J, H, and K$_{s}$ are 250.0, 187.0, and 247.0 minutes, respectively. Dark frames were also obtained for each integration time.

The same basic reduction technique was applied to all three bands using the pipeline of Tom Jarret~\footnote{http://spider.ipac.caltech.edu/staff/jarrett/wirc/jarrett.html}. A median dark frame was subtracted from the data frames, and correction terms for the flux non-linearity are applied. A median sky made of a maximum of 10 frames was calculated and subtracted from the data frame before flat fielding. By accounting for the rotational offset of the telescope using a list of known stars from the Two Micron All Sky Survey (Skrutskie et al. 2006; hereafter, 2MASS) the astrometry was measured and verified by eye. The background level and flux bias corrections were then subtracted from the frames.  {\it SWarp} \citep{ber07} was used to mosaic the frames together. A final flux calibration on the mosaic, relative to the 2MASS catalog, was performed.

\nocite{skr06}

\begin{figure*}
\epsscale{1.0}
\plotone{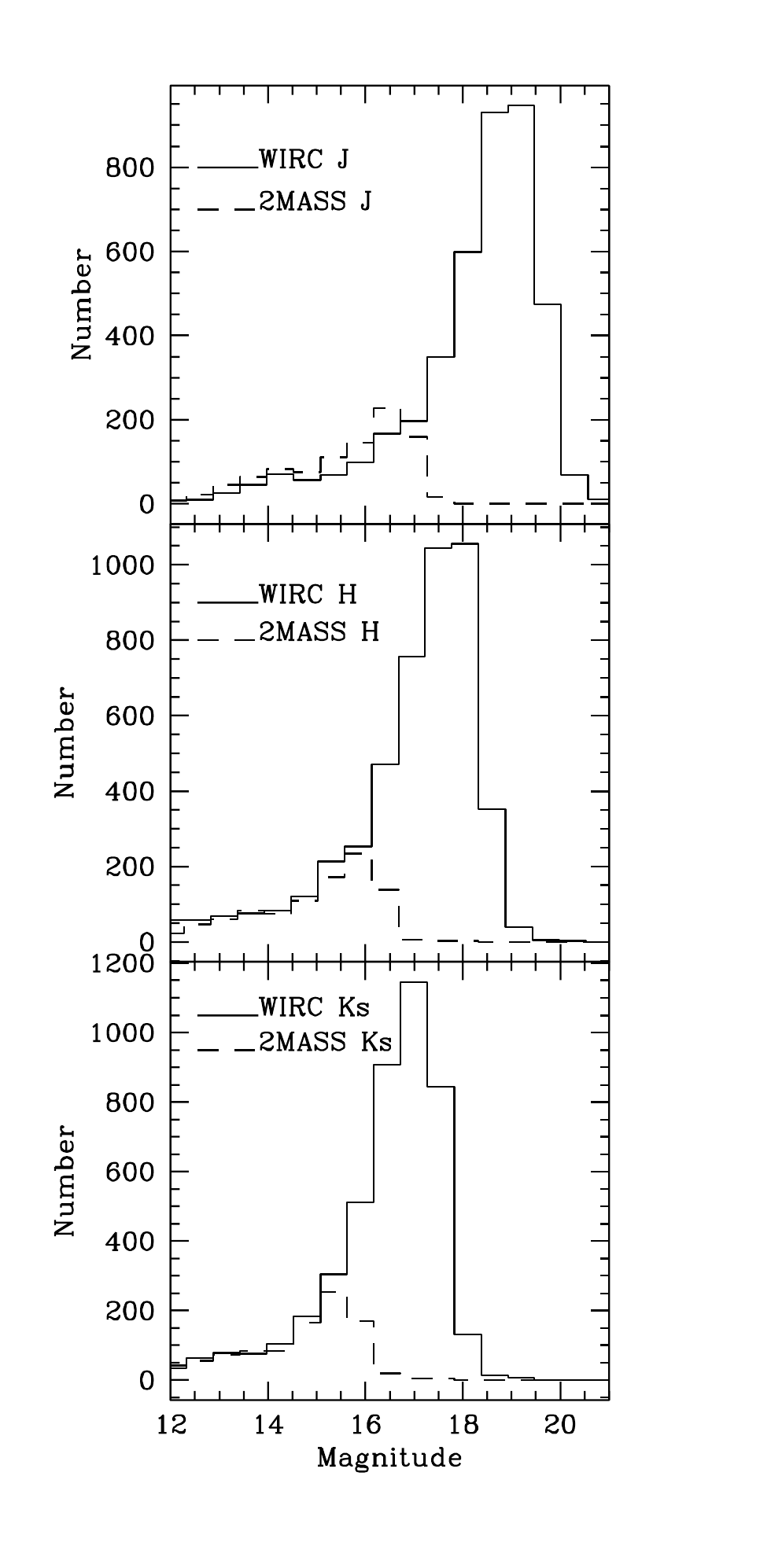}
\caption{{\bf Depth of Near-IR data.}  Our new Near-IR images reach a depth of approximately two magnitudes deeper than those available from 2MASS.
\label{jhk2mass}
}
\end{figure*}

The Near-IR photometry reaches a depth of J$\sim$19.5, H$\sim$18.5, and K$_{s}\sim$ 17.0 with uncertainties of 1.9, 2.7, and 4.6\%, respectively. Figure~\ref{jhk2mass} demonstrates how this is roughly two magnitudes deeper than those available with 2MASS.

%\begin{figure}
%\epsscale{0.8}
%\plotone{JKcol}
%\caption{{\bf Near-IR color.}  Points show the J-K$_{s}$ color for all galaxies in the field.  Members are shown as black circles, and members within 0.35$\,$Mpc of the cluster center are shown as grey circles.
%\label{jkcol}
%}
%\end{figure}

SExtractor \citep{ber96} was used to identify sources and measure their magnitudes. To ensure we extracted both small and  extended objects, we used a {\em det\_minarea} of 5 and {\em det\_minthresh} of 2.0. To ensure we found the smaller objects that often surround larger objects (ex. near the brightest cluster galaxy) we used a {\em deblend\_mincont} of 0.00005.

%Figure~\ref{jkcol} shows that the Near-IR color-magnitude relation has a tight red-sequence. The Near-IR (J$-$K$_{s}$) color is driven by old stars, which comprise the bulk of the galaxy mass, and therefore are paramount in our later derivation of the galaxy stellar mass.

\subsection{{\em Spitzer} Observations}

We have collected archival photometry from the {\it Spitzer} Space Telescope (PIs: Fazio, Rieke, Kitayama). The data were taken in 2003, 2004, and 2007 and all sets were observed for at least 1.8 hours (see Table~\ref{obs}). The MIPS instrument provides Far-Infrared images at 24$\mu$m, 70$\mu$m and 160$\mu$m. IRAC data cover the Mid-Infrared wavelengths of 3.6, 4.5, 5.8, and 8.0$\mu$m, in Figure~\ref{spitzer_cov} they are illustrated as dashed lines and available only for the cluster core.

Each photometric band is analyzed on its own to produce a catalog of galaxies.  Dust enshrouded bright stars re-emit their energy at longer wavelengths, and the 24$\mu$m flux approaches this peak of the spectral energy distribution (SED), allowing for an estimate of star formation rates.  Although MIPS 70$\mu$m and 160$\mu$m observations go even further toward the SED peak, these detectors are not as sensitive as those at 24$\mu$m. Therefore, we choose to base our catalogs on the 24$\mu$m sources. Sources from the 24$\mu$m catalog are then matched to candidates from each of the other photometric bands (see Section \ref{24cat} for details). The main steps in data processing are outlined below. More detailed descriptions can be found in \citet{edw10} and \citet{fad06}.

\subsubsection{MIPS 24$\mu$m Data Reduction}

The Basic Calibrated Data (BCDs) and calibration files are downloaded from Leopard, pipeline version S17.0.4, which uses a factor of  0.045 to convert instrumental units to MJy~sr$^{-1}$. These BCDs have already been corrected for dark subtraction, muxbleed, detector linearization, flat fielding, and cosmic rays. We redid the pipeline jailbar correction after masking bright sources, applied a droop and superflat correction for all BCDs in each AOR, and apply a flat for each position of the cryogenic scan mirror. We removed the zodiacal light and corrected the astrometry to match the SDSS source positions.

The BCDs were mosaiced using MOPEX (\citet{mako05,mak05}) and extended sources were extracted using SExtractor. We ran SExtractor a second time on the residual image, after multiplying by 141.08 to convert MJy~sr$^{-1}$ to $\mu$Jy (as listed in the MIPS Data Handbook~\footnote{http://ssc.spitzer.caltech.edu/mips/dh/}, p. 31). We computed aperture fluxes within a radius of 10.3$^{\prime\prime}$ as well as Petrosian magnitudes with 4 radii and a minimum radius of 5.5$^{\prime\prime}$. We applied a factor of 1.167 for the aperture correction~\footnote{http://ssc.spitzer.caltech.edu/mips/apercorr/} and divided by 0.961 to account for the color correction as listed in the MIPS Data Handbook.  The data were combined from two different programs, leaving the central region much deeper. Although we included this deep strip in the reduced images, we note that the bulk of our analysis is based on a spectroscopic sample which corresponds only to the depth of the full MIPS image. By measuring the noise in the mosaiced frames, we arrived at a 5$\sigma$ depth of 80$\,$$\mu$Jy in the 24$\mu$m mosaic. The final catalog includes sources with a signal to noise ratio (SNR) $>$3 sources, computed using an optimal aperture radius.

There is one source, NGC~4921, which is extremely extended and has many star forming clumps at 24$\mu$m. SExtractor failed for this source, therefore, we calculated the photometry by hand, fitting a Petrosian radius to the portion of the light profile which encompasses the same region as seen at r$^{\prime}$.

\subsubsection{MIPS 70$\mu$m and 160$\mu$m Data Reduction}

For the longer wavelength MIPS observations, we downloaded the raw frames (DCEs) from the {\em Spitzer} archive and use the GeRT~\footnote{http://ssc.spitzer.caltech.edu/mips/gert/} pipeline which finds and calibrates the slope of consecutive data reads.  The conversion factor for instrumental units into MJy~sr$^{-1}$ is 702 for the 70$\mu$m data, and 47.4, for the 160$\mu$m data. For the 70$\mu$m data, we used a second pass filtering technique to remove streaking in the final image. We identified the causal bright sources in a first pass, and repeat the filtering on the masked image. We applied an additional correction to the BCDs to remove latent stim artifacts by modelling and then removing the effect of the stimflash on the median of the BCDs that follow. 

Virtually all of the extragalactic sources at these longer wavelengths are point sources. We mosaiced the frames and conducted the extraction of the sources using {\em APEX} with a bottom outlier threshold of 2 and a top threshold of 3. For the 70$\mu$m data, we used an aperture radius of 15$^{\prime\prime}$, a color correction of 0.918, and an aperture correction of 2.244. For 160$\mu$m, we use an aperture radius of 20$^{\prime\prime}$, a color correction of 0.959, and an aperture correction of 3.124. By measuring the noise in the mosaiced frames, we arrived at a 5$\sigma$ depth of 32.0$\,$mJy in the 70$\mu$m mosaic and 130$\,$mJy in the 160$\mu$m mosaic.

\subsubsection{IRAC Data Reduction}

For all four IRAC bands, we included corrections for the column pulldown, jailbars, zodiacal light, subtraction of bright stars and background level, as well as a treatment for the background droop near bright stars.  The only alteration to the process discussed in detail in \citet{edw10} is that for IRAC 5.8 and 8.0$\mu$m, we included a correction for the superflat before subtracting the bright stars, this removes the gradient inside each BCD.

The IRAC data reach a uniform 5$\sigma$ depth of 1.5, 4.0, 3.9, and 3.8$\mu$Jy in channels 1-4, respectively. As with the MIPS data, there is a deeper patch of 137$\,$arcmin$^{2}$ in each channel. Although the area of the deeper observations is consistent in all channels, the center of the deeper data is not consistent. Nonetheless, our spectroscopic photometric cut prevents the use of the deeper central data strip.

The IRAC sources were extracted using SExtractor and sources with close pairs (within a projected separation $<$7$^{\prime\prime}$) were flagged.  For IRAC~3.6$\mu$m, ~4.5$\mu$m, IRAC~5.8$\mu$m, and IRAC~8.0$\mu$m, 11, 10, 7, and 9\%, respectively, of the sources are close pairs. If an IRAC source was within 1$^{\prime\prime}$ of a star as determined by the SDSS star/galaxy separator flag, it was flagged as a star. For IRAC ~3.6$\mu$m, ~4.5$\mu$m, IRAC~5.8$\mu$m, and IRAC~8.0$\mu$m, 6, 7, 7, and 6\% of the sources were flagged as stars, respectively.

We calculated aperture fluxes using diameters of 4,6, and 12$^{\prime\prime}$.  The aperture corrections we used follow those of the SWIRE technical note \citep{sur05} which lists corrections for radii of 1.9, 2.9 and 5.8$^{\prime\prime}$. Following their work, we divided the aperture fluxes in each band by the corrections at each aperture radius. The corrections for each aperture radius are 0.736, 0.87, and 0.96, for IRAC~3.6$\mu$m; 0.716, 0.87, and 0.95, for IRAC~4.5$\mu$m; 0.606, 0.80, and 0.94, for IRAC~5.8$\mu$m; and 0.543, 0.700, and 0.940 for IRAC~8.0$\mu$m. To estimate the total flux we also quote Petrosian fluxes. These are based on apertures whose radius is the average of the azimuthal light profile out to the radius where the local surface brightness is a factor of 0.2 times the mean surface brightness within the radius \citep{pet76}. We used two Petrosian radii and a minimum radius of 3$^{\prime\prime}$, and applied the aperture correction for extended sources as explained on the IRAC calibration webpage~\footnote{http://ssc.spitzer.caltech.edu/irac/calib/}. We quote the uncertainty of the fluxes as those given by SExtractor, known to be underestimates as they only represent the error from Poissonian noise \citep{val09}. Therefore, we also include maps of the image root-mean square (RMS), calculated from the image and coverage maps as described in Section~\ref{Images}.

\subsection{Optical Photometry and Spectroscopy}

Using the SDSS Catalog Archive Server, we collect the photometric information for the 97660 sources available with data release 7 (DR7) which overlap with the 24$\,\mu$m field of view. This data release covers imaging in all five optical bands, u$^{\prime}$,g$^{\prime}$,r$^{\prime}$,i$^{\prime}$, and z$^{\prime}$ down to r$^{\prime}$$<$22.2 at a resolution of 1.4$^{\prime}$$^{\prime}$ in r$^{\prime}$.  After matching to our Mid-IR catalog of 7603 sources, we find that 4469 of the 24$\mu$m galaxies have Sloan galaxy counterparts.

We collect the Petrosian magnitudes from Sloan, again, replacing NGC~4921 with our own measurements, although in this case, the two values are similar.

\subsubsection{Archival Spectra}

Inside the MIPS field of view, there are 346 galaxies with SDSS spectra and redshifts between 0.015 and 0.035. These spectra are observed with an aperture diameter of 3$^{\prime}$$^{\prime}$, cover the wavelength range from 3800 to 9200$\,$\AA, and have a spectral resolution of 3$\,$\AA~in r$^{\prime}$. Redshifts measured from these spectra are accurate to $\sim$30$\,$km/s. 

We also gather spectroscopic redshifts for 869 members from publicly available catalogs in NED and elsewhere in the literature.

\subsubsection{WIYN Spectra}

\begin{figure*}
\epsscale{1.0}
\plotone{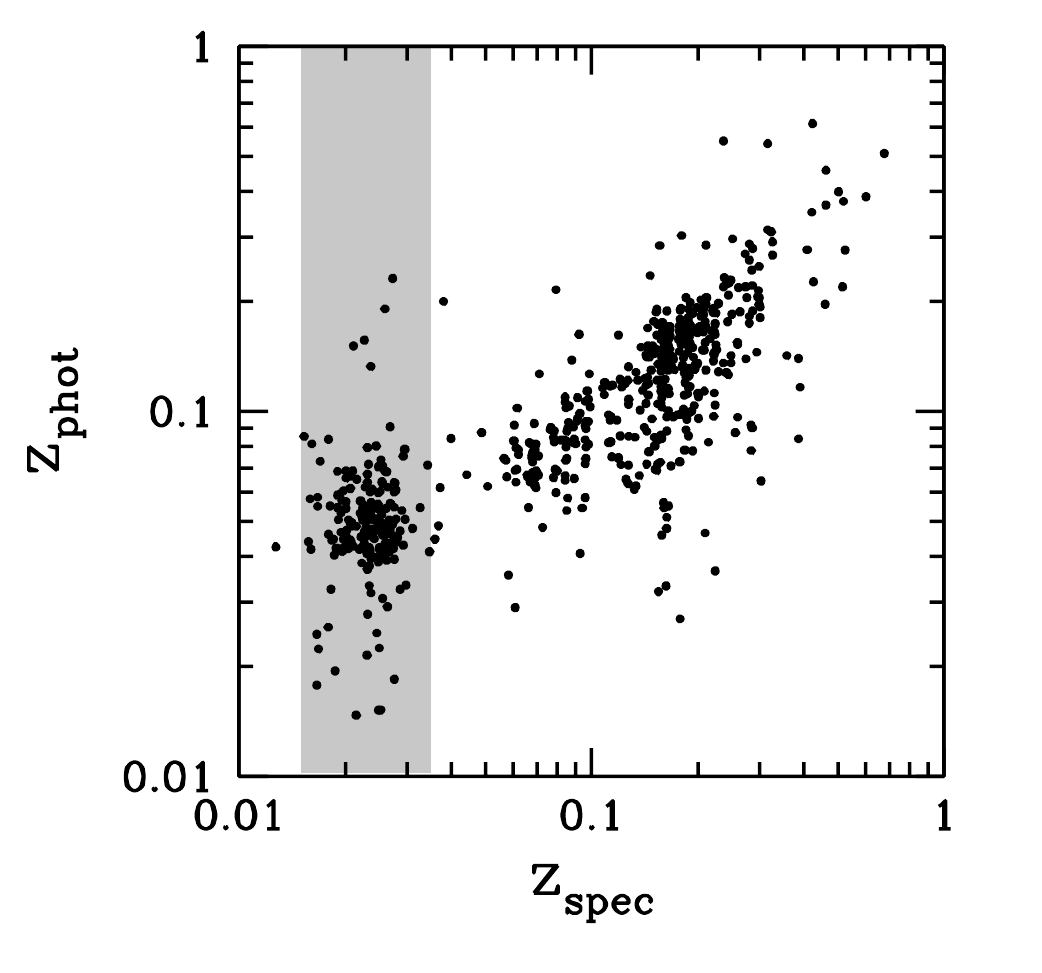}
\caption{{\bf Photometric redshift cut.} The reliability of the photometric redshift estimates: spectroscopically confirmed cluster members (shaded region) have z$_{spec}$ between 0.015 and 0.035 and most have z$_{phot}$ $<$ 0.08. \label{pickr}}
\end{figure*}

To add to the Sloan spectra, we include our own spectroscopic observations targeting 24$\mu$m sources. We were awarded three nights with the Hydra multiobject spectrometer on WIYN in July, 2008, however lost one night due to problems with telescope focusing. We observed 338 out of 529 proposed sources which have magnitudes of r$^{\prime}$ $<$ 19.5, photometric redshifts of $z_{phot}< $0.2, and 24$\,\mu$m counterparts. By comparing to galaxies with photometric redshifts $<$$\,$0.08 (see Figure~\ref{pickr}), we determine our spectroscopic sample to be 80\% complete down to r$^{\prime}$ $<$ 19.5, although the completeness drops to 60\% in the faintest bin (18.5 $<$ r$^{\prime}$ $<$ 19.5). The restriction in photometric redshift helps target a higher fraction of true spectroscopic cluster members, as can be seen in Figure~\ref{pickr}. 

To obtain spectra between 3990-6795$\,$\AA~we integrated for 1200$\,$s with the blue cable and  grating 600@10.1, centered at 5380$\,$\AA. This gives a dispersion of 1.4$\,$\AA/pixel, a spectral coverage of 2850~\AA, and a resolution of 4.6$\,$\AA.

The spectra were reduced using IDL scripts that trim and divide the science image by average dome flats, subtract the bias, perform calibration of the lines using arc lamps, and calibration of the fluxes using filtered standard stars. Cosmic rays were removed using the {\it la\_cosmic} \citep{van01} task. The scripts were run for each configuration and for each night. More details on the reduction methods can be found in \citet{mar07}.

Seventy-six of the new spectra belong to cluster member galaxies with spectroscopic redshifts between 0.015 and 0.035. Although many of these members already have published redshifts in the literature, with these spectra we increase the number of member galaxies for which we can perform emission line diagnostics by 20\%. Additionally, because we have targeted the MIPS 24$\mu$m emitting galaxies in our spectroscopic survey, we increase the number of known infrared members to 210. We gathered our spectral data in 2008, based on the SDSS DR6, and SDSS DR7 has since covered many of the members we observed. Still, our WIYN spectra increase the number of infrared spectral sources by 13\%.

\subsubsection{Line Measurements}

Optical spectra are useful threefold: First, they determine cluster membership; second, they allow for a measure of the unobscured star formation rate based on emission lines; and third, the emission line ratios are a powerful diagnostic of the nature of line emission (either from Active Galactic Nuclei (AGN) or star formation).

We use a custom IDL program to measure the emission and absorption lines for both the WIYN and SDSS spectra. The code makes use of the Markwardt package curve-fitting algorithms. The galaxy redshifts are determined by cross-correlating SDSS galaxy templates, the continuum is fit by hand, and the emission and absorption features are fit concurrently, using Gaussian functions. Close lines, such as the [NII]-H$\alpha$ triplet, are deblended.

The line measurements for the 87 MIPS emission line galaxies, for which we can access full spectroscopic information, are given in Table~\ref{elines}. The full table is available in the electronic version of this paper, though we present the first five lines in both versions.

\begin{deluxetable}{lcccccccccc}
\tabletypesize{\scriptsize}
\tablewidth{0pt}
%\tablecolumns{<num>}
\tablecaption{Emission Line Measurements for Member Galaxies \label{elines}}
\tablehead{\colhead{ID} & \colhead{RA} & \colhead{DEC} & \colhead{z} & \colhead{SFR$_{IR}$} & \colhead{Mass} & \colhead{SFR$_{H\alpha}$} & \colhead{H$\beta$}& \colhead{[OIII]} &\colhead{H$\alpha$}  & \colhead{[NII]}\\
 &\multicolumn{1}{c}{(J2000)} & \multicolumn{1}{c}{(J2000)} & & \multicolumn{1}{c}{(M$_{\odot}$/yr)} & \multicolumn{1}{c}{(10$^{9}$ M$_{\odot}$)} & \multicolumn{1}{c}{(M$_{\odot}$/yr)} & \multicolumn{4}{c}{(10$^{-17}$erg$\,$s$^{-1}$cm$^{-2}$)} 
}
\startdata
  1 & 194.757507 & 26.815569 & 0.0236 & 0.06 & 0.47 & - &   - &    - &    - &   - \\ 
  2 & 194.976395 & 26.820110 & 0.0268 & 0.01 & 3.78 & 0.17 &  11 $\pm$   4 &   5 $\pm$   3 & 124 $\pm$   5 &  59 $\pm$   4 \\ 
  3 & 195.014603 & 26.898619 & 0.0196 & 0.05 & 31.54 & - &    -&    -&    -&    -\\ 
  4 & 194.936707 & 26.954750 & 0.0343 & 0.33 & 2.80 & 0.21 & 107 $\pm$   4 &  43 $\pm$   3 & 487 $\pm$  11 & 152 $\pm$   7 \\ 
  5 & 194.955200 & 26.974531 & 0.0234 & 0.26 & 0.57 & - & 116 $\pm$   8 &  77 $\pm$   7 & 374 $\pm$   9 &  81 $\pm$   5 \\ 
\enddata
\end{deluxetable}

\section{Images, Spectra, and Catalogs}\label{Images}

We provide all of our images and photometric catalogs to the community online through {\em IRSA}~\footnote{http://irsa.ipac.caltech.edu}. 

The fully reduced and calibrated WIRC J, H, and K$_{s}$ images are in instrumental units. IRAC 3.6, 4.5, 5.8, and 8.0~$\mu$m, and MIPS 24, 70, and 160$\,\mu$m images are in units of MJy~sr$^{-1}$. Coverage, uncertainty, weight, and RMS maps are included where available. The coverage maps are in units of number of BCDs. The uncertainty maps, which can be used to understand the errors, are computed from MOPEX accounting for Poissonian noise as well as a Gaussian component from the read out. The weight maps are computed in {\it SWarp} and are maps of the local sky background flux variance. RMS maps are computed on a pixel to pixel basis: the median value of the coverage map is divided by the value of the coverage map at each pixel, then the square root of this ratio is multiplied by the standard deviation of the image. All images discussed below are in FITS format.

{\em WIRC} - Included are intensity and weight maps. All images have a pixel size of 0.249$^{\prime\prime}$. The image sizes are 4.3~Gbytes.

{\em IRAC} - Intensity, coverage, and uncertainty maps are provided. All images are 206~Mbytes and have a pixel size of 1.27$^{\prime\prime}$.

{\em MIPS} - Intensity, coverage, and uncertainty maps are included. For MIPS 24$\mu$m the image sizes are 173~Mbytes and the pixel size is 1.27$^{\prime\prime}$. For MIPS 70$\mu$m the image sizes are 18~Mbytes and the pixel size is 4.00$^{\prime\prime}$. For MIPS 160$\mu$m the image sizes are 2.4~Mbytes and the pixel size is 8.00$^{\prime\prime}$.

We also provide the WIYN spectra we observed with WIYN in the region of Coma, to the NED spectral database.

\subsection{24$\mu$m Catalog}\label{24cat}

The master catalog is based on the MIPS 24$\mu$m source catalog. It is constructed by finding the best counterpart to each MIPS source in each of the other bands: u$^{\prime}$, g$^{\prime}$, r$^{\prime}$, i$^{\prime}$, and z$^{\prime}$, J, H and K$_{s}$ magnitudes, and, flux densities from IRAC (3.6, 4.5, 5.8 and 8.0$\,$$\mu$m) and MIPS 70$\mu$m and 160$\mu$m.

To discern real counterparts to the 24$\mu$m sources from background sources, we use the methods of \citet{sut92} and \citet{cil03}. Briefly, this involves choosing the candidate with the highest reliability.  The reliability is the likelihood ratio normalized over all possible candidates and accounts for the separation as well as the relative magnitude distribution of the two data sets (24$\mu$m and counterpart). We consider only associations within a 5$^{\prime\prime}$ radius of the MIPS 24$\mu$m source, and which have a likelihood ratio greater than 0.2, see \citet{fad06} for a more detailed description. The 70$\mu$m and 160$\mu$m images have lower resolution and we simply choose the candidate with the shortest projected distance. If a 24$\mu$m source does not have match in a particular band, it is flagged with -999. 

\begin{deluxetable}{lcl}
\tabletypesize{\scriptsize}
\tablewidth{0pt}
%\tablecolumns{<num>}
\tablecaption{MIPS 24 $\mu$m Source Catalog Columns \label{cat24} }
\tablehead{\colhead{Column} & \colhead{Format} & \colhead{Description}}
\startdata
1 & i6 & Catalog Number\\
2 &f14.6 &RA (J2000)\\
3 &f14.6 &DEC (J2000)\\
4 &f10.3 &MIPS24 ($\mu$Jy)\\
5 &f10.3 &MIPS24 error ($\mu$Jy)\\
6  &f9.3 & SNR \\
7  &f9.3 & MIPS24-SDSS Sep ($^{\prime\prime}$)\\
8  &4xf9.3 & MIPS24-IRAC Sep($^{\prime\prime}$)  \\
12  &3xf9.3 & MIPS24-NIR Sep($^{\prime\prime}$) \\
15  &f9.3 & MIPS24-70 Sep($^{\prime\prime}$)\\
16  &f9.3 & MIPS24-160 Sep($^{\prime\prime}$)\\
17  &f9.3 & MIPS24-z Sep($^{\prime\prime}$)\\
18 &f10.3 &u$^{\prime}$ (mag)\\
19 &f10.3 &g$^{\prime}$ (mag)\\
20 &f10.3 &r$^{\prime}$ (mag)\\
21 & f10.3 &i$^{\prime}$ (mag)\\
22 &f10.3 &z$^{\prime}$ (mag)\\
23 & 2xf10.3 &J ap 3.5$^{\prime\prime}$,Petrosian (mag)\\
25 & 2xf10.3 &H ap 3.5$^{\prime\prime}$,Petrosian (mag)\\
27 & 2xf10.3 &K$_{s}$ ap 3.5$^{\prime\prime}$,Petrosian (mag)\\
29 & 4xf10.2 &IRAC1 ap 4$^{\prime\prime}$,ap 6$^{\prime\prime}$,ap 12$^{\prime\prime}$,Petrosian ($\mu$Jy)\\
33 & 4xf10.2 &IRAC2 ap 4$^{\prime\prime}$,ap 6$^{\prime\prime}$,ap 12$^{\prime\prime}$,Petrosian ($\mu$Jy)\\
37 & 4xf10.2 &IRAC3 ap 4$^{\prime\prime}$,ap 6$^{\prime\prime}$,ap 12$^{\prime\prime}$,Petrosian ($\mu$Jy)\\
41 & 4xf10.2 &IRAC4 ap 4$^{\prime\prime}$,ap 6$^{\prime\prime}$,ap 12$^{\prime\prime}$,Petrosian ($\mu$Jy)\\
45 &f10.2 &MIPS70 ($\mu$Jy)\\
46 &f10.2 &MIPS160 ($\mu$Jy)\\
47 &f10.3 &u$^{\prime}$ error (mag)\\
48 &f10.3 &g$^{\prime}$ error (mag)\\
49 &f10.3 &r$^{\prime}$ error (mag)\\
50& f10.3 &i$^{\prime}$ error (mag)\\
51 &f10.3 &z$^{\prime}$ error (mag)\\
52 & 2xf10.3 &J error ap 3.5$^{\prime\prime}$,Petrosian (mag)\\
54 & 2xf10.3 &H error ap 3.5$^{\prime\prime}$,Petrosian (mag)\\
56 & 2xf10.3 &K$_{s}$ error ap 3.5$^{\prime\prime}$,Petrosian (mag)\\
58 & 4xf10.2 &IRAC1,IRAC1 error ap 4$^{\prime\prime}$,ap 6$^{\prime\prime}$,ap 12$^{\prime\prime}$,Petrosian ($\mu$Jy)\\
62 & 4xf10.2 &IRAC2,IRAC2 error ap 4$^{\prime\prime}$,ap 6$^{\prime\prime}$,ap 12$^{\prime\prime}$,Petrosian ($\mu$Jy)\\
66 & 4xf10.2 &IRAC3,IRAC3 error ap 4$^{\prime\prime}$,ap 6$^{\prime\prime}$,ap 12$^{\prime\prime}$,Petrosian ($\mu$Jy)\\
70 & 4xf10.2 &IRAC4,IRAC4 error ap 4$^{\prime\prime}$,ap 6$^{\prime\prime}$,ap 12$^{\prime\prime}$,Petrosian ($\mu$Jy)\\
74 &f10.2 &MIPS70 error ($\mu$Jy)\\
75 &f10.2 &MIPS160 error($\mu$Jy)\\
76 &f10.2 &SDSS Photz \\
77 &f10.2 &SDSS Photz error \\
78 &f10.2 &Spec z\\
79 &f10.2 &Spec z source 1:LIT,3:SDSS,4:WIYN,5:NED\\
80&f9.3 & MIPS24-z Sep($^{\prime\prime}$)\\
81 &i2 & MIPS24 separation flag \\
82 &i2 & MIPS24 star/galaxy flag \\
83& 7xf7.3 & IRAC1-4,J,H,K stellarity index\\
90& 7xi2 & IRAC1-4,J,H,K close pair flag\\
97& 7xi2 & IRAC1-4,J,H,K star/galaxy flag\\
\enddata
\end{deluxetable}

Table~\ref{cat24} describes the machine readable 24$\mu$m source catalog. The first column lists the columns of the catalog that correspond to each entry. The second column lists the format of those columns, whether it is an integer or float, and the number of characters. The third column gives a short description of each entry. Information in the table includes the catalog number, the J2000 RA and DEC position in decimal degrees, followed by the flux and error of the MIPS 24$\mu$m source in $\mu$Jy and its SNR. Next, we list the separation between the 24$\mu$m source and its separation from each candidate. The SDSS u$^{\prime}$, g$^{\prime}$, r$^{\prime}$,i$^{\prime}$,z$^{\prime}$ Petrosian magnitudes are listed, followed by the WIRC J, H, and K$_{s}$ aperture and Petrosian magnitudes. The SDSS magnitudes are given in their native AB magnitude system. The J, H and K$_{s}$ magnitudes are in the Vega magnitude system. We list three aperture fluxes and the Petrosian flux for the four IRAC bands, and the PSF flux for MIPS 70$\mu$m and 160$\mu$m. We then list the photometric errors, respectively. This is followed by the SDSS photometric redshift, its error, and the spectroscopic redshift. Finally several flags are listed. The first gives the source of the spectroscopic redshift. Values gathered from the literature are marked with a `1', SDSS values with a `3', WIYN values with `4', and NED values with `5'. The next two are given a value of 1 for true; the first flags close pairs of  24$\mu$m sources (those within 6.1$^{\prime\prime}$ of another source, we note that less than one percent have close pairs).  The second flag marks the MIPS 24$\mu$m sources associated with stars as defined by the SDSS star/galaxy separator. The next seven values list the stellarity index of, respectively, the IRAC 3.6 - 8.0$\mu$m sources, followed by the WIRC J, H, K$_{s}$ sources (all calculated using SExtractor). The final 14 flags are the close pair flags for the IRAC and WIRC bands followed by the star/galaxy flags for the IRAC and WIRC bands.

\section{Results}

Our aim is to derive the specific star formation rate (sSFR) for Coma galaxies by taking advantage of the large baseline of multi-wavelength data available. As a first step, we classify the galaxy activity as either star formation, or as contaminated by AGN. We employ several diagnostics, including optical emission line ratios, IRAC colors, and the Radio Far-Infrared (radio-FIR) correlation. Next, we build the SEDs based on up to 15 photometric datapoints and match these to empirical template galaxy spectra from which we measure the total infrared luminosity, and galaxy stellar mass. We find that galaxies with specific SFRs $>\,$0.1$\,$Gyr$^{-1}$ are blue, and that the majority are dwarfs, having mass$\,$$<$$\,$10$^{9}$M$_{\odot}$.

\subsection{AGN Contamination}\label{agnsec}

AGN activity can mimic many of the same emission characteristics as young starbursts, including bright MIPS detections. Using several additional diagnostics, we show that the contamination from AGN is low. 

\subsubsection{Optical Emission Lines}

\begin{figure*}
\epsscale{1.0}
\plotone{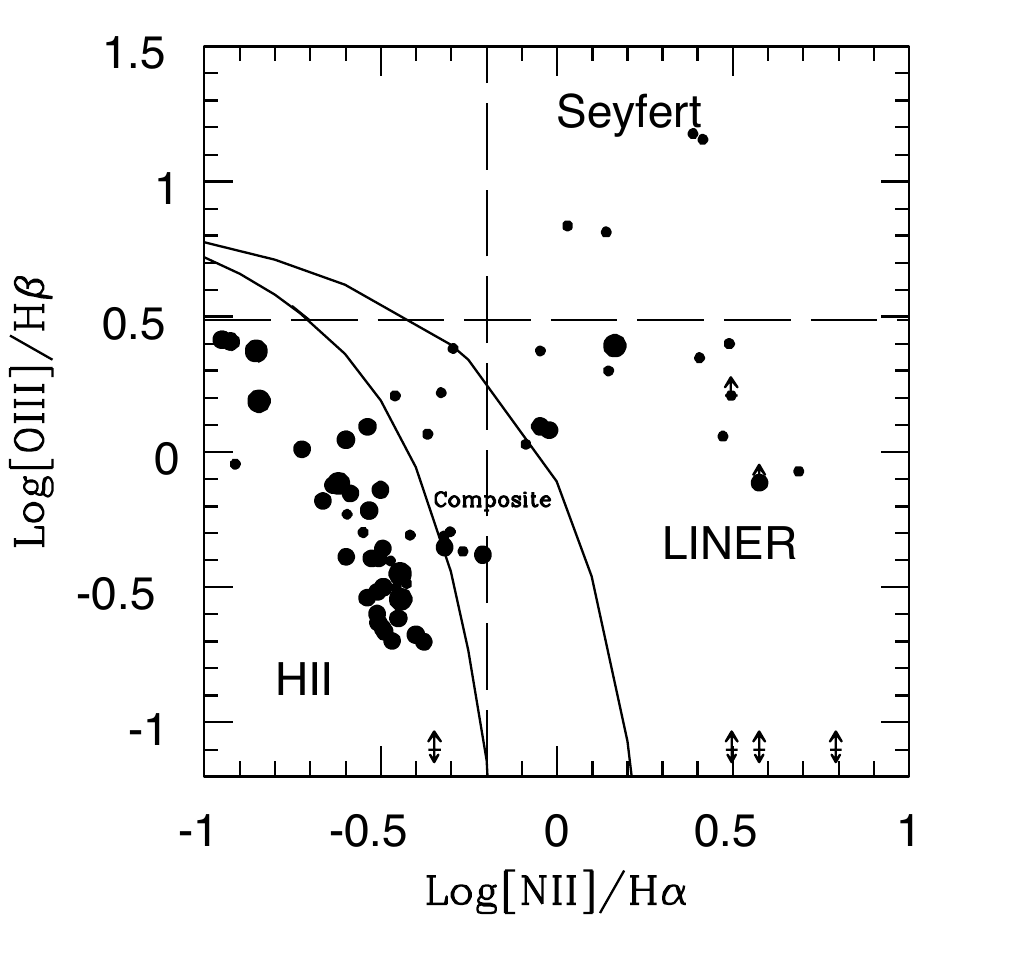}
\caption{{\bf BPT diagram.} Ratios of nebular emission lines can be used to separate optical AGN from clear star forming galaxies. The dashed lines separate the star forming and AGN regions of \citet{ost06} and the curved lines separate  star forming, composite, and AGN sources \citep{kau03}. Plotted, are the cluster members which contain measurable line emission. The majority fall in the starburst and transition regions of the diagnostic diagram, with few clear AGN. Larger dots represent higher specific star formation rates, based on SED fitting, and assuming all the luminosity comes from star formation (see the discussion in Section~\ref{ssfrsection4}. }\label{bpt}
\end{figure*}

Of the 210 known cluster members with MIPS 24$\mu$m emission, 125 have spectra. Of these last, 38 are passive galaxies with no measurable emission lines. Table~\ref{elines} presents the emission line measurements of the remaining 87 used to construct the BPT diagram \citep{bpt81} of Figure~\ref{bpt}. The dashed lines separate the star forming and AGN regions of \citet{ost06}, and the curved lines show separations between star forming galaxies from the SDSS \citep{kau03}, composite sources, and a theoretical upper limit for sources that can be described by star forming models \citep{kew02}. Only four are Seyfert galaxies, and 14 LINER. Most emission line galaxies fall in the star forming region, or in the transition region where star formation is still a dominant process.

\subsubsection{IRAC Colors}

\begin{figure*}
\epsscale{1.0}
\plotone{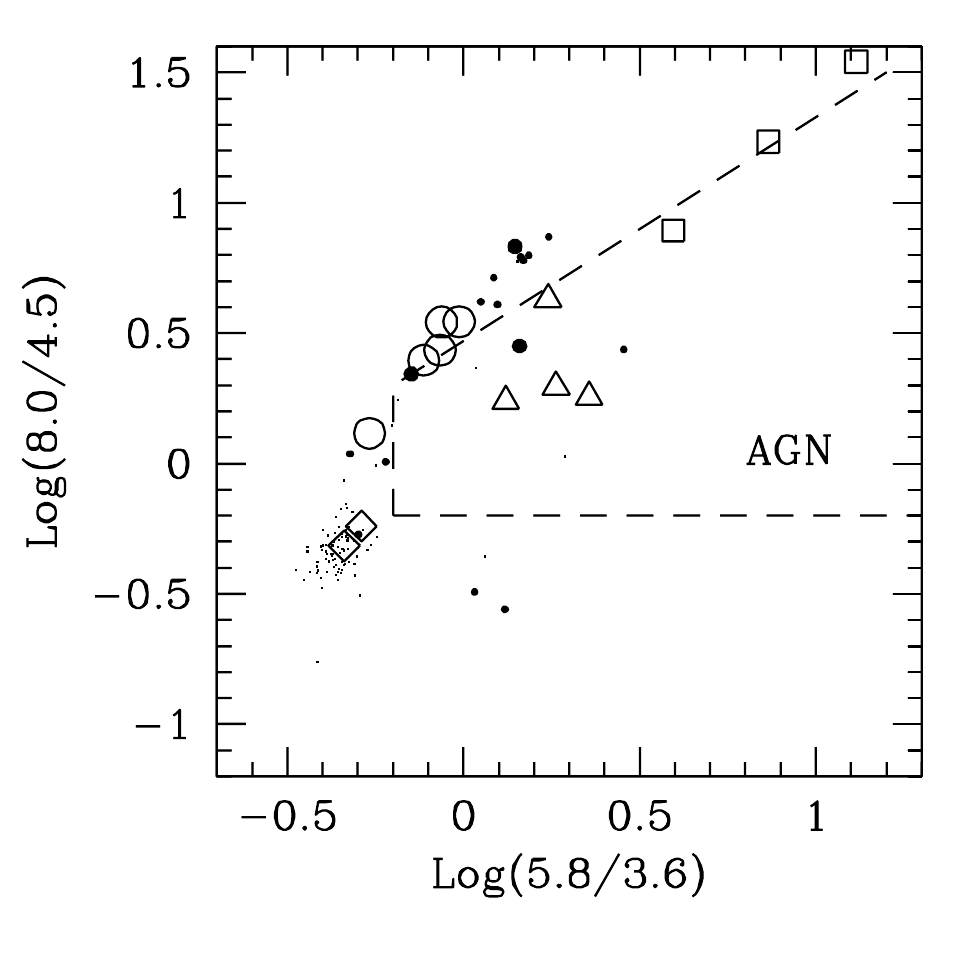}
\caption{{\bf IRAC colors.} Stars, ellipticals, and spirals can be separated using IRAC colors. We plot MIPS selected members which have detections in all four IRAC bands, following \citet{lac04}. Black dots are for passive galaxies, small filled dots are low and star forming galaxies, and the larger filled dots are starbursts. Open shapes represent the IRAC colors from several templates: ellipticals (diamonds), spirals (circles), AGN (triangles), starbursts (squares). Very few of the spectroscopically confirmed members are inside of the dashed polygon, the AGN region of the plot.}\label{iracAGN}\label{iracAGNmod}
\end{figure*}

Stars, galaxies, and broad line AGN can be differentiated using their IRAC colors \citep{eis04,lac04,saj05,don08}. AGN are well separated because the power law of AGN emission is more red than the galaxy spectrum in the 3.6$\mu$m channel, and because there is a lack of PAHs in AGN. \citet{lac04} plot the IRAC colors as (8.0/4.5) vs (5.8/3.6) which separates objects with blue continua from objects with red continua; AGN tend to be red in both IRAC color filters.

We plot the IRAC colors of our MIPS selected galaxies in Figure~\ref{iracAGN}. The majority of bright IR sources avoid the AGN region, although four may be classified as AGN.  The predicted IRAC colors from template galaxies used in our SED fitting, at the redshift interval of the Coma Cluster, are also shown in the figure. As can be seen, the template AGN (triangles) occupy approximately the same locus as defined by \citet{lac04}.

\subsubsection{Radio Far-Infrared Correlation}

  \begin{figure*}
   \center
  \epsscale{1.0}
     \plotone{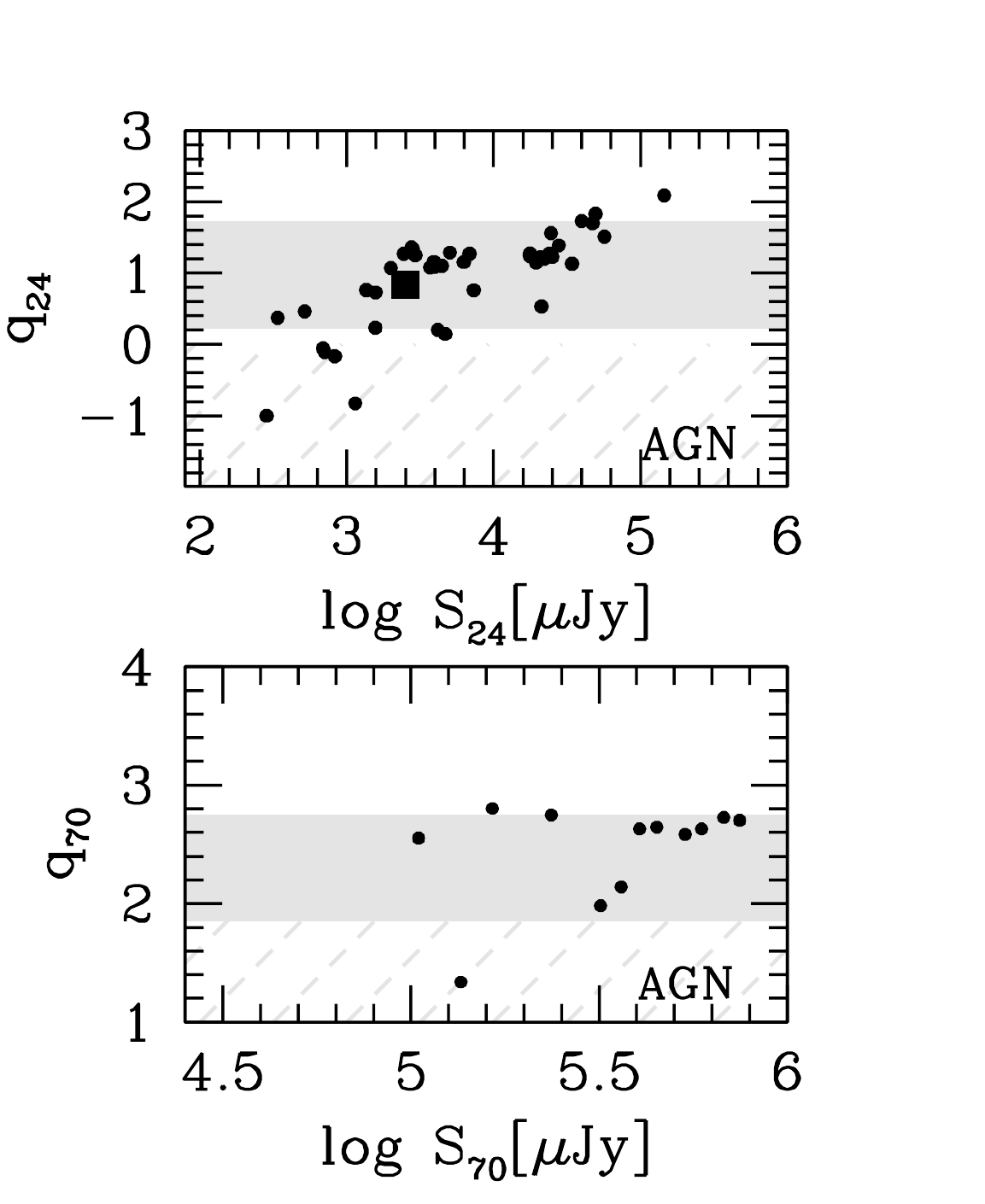}
    \caption{{\bf The radio-FIR correlation.} The plots show the logarithm of the observed flux density ratio (q) versus the infrared flux density for all spectroscopic members with both a MIPS and radio detection. Sources with MIPS 24$\mu$m emission are on the top, and with MIPS 70 $\mu$m emission are on the bottom. The solid grey bar indicates the best-fit radio-FIR correlation and its 3$\sigma$ error.  The grey hashed area under the correlation band is expected to be populated only by radio-excess AGN.\label{qval}}
\end{figure*}

Using the radio-FIR correlation for star forming galaxies of \citet{con92}, we can determine the presence of radio-excess AGN. We use the catalog of 1.4GHz radio fluxes for Coma galaxies from \citet{mil09} and match the radio galaxies to our MIPS 24$\mu$m and 70$\mu$m sources. The 1.4GHz resolution is similar to that of the 24$\mu$m resolution, and we match our sources purely on proximity in RA, Dec, and redshift. As shown in Figure~\ref{qval}, sources well below the relationship harbor a radio excess, this is attributed to AGN emission. From \citet{app04}, the limiting ratios are q$_{24}$= log~(S$_{24\mu m}$/S$_{1.4GHz}$)$\sim$ 1$\,\pm\,$0.27 and q$_{70}$= log~(S$_{70\mu m}$/S$_{1.4GHz}$) $\sim$ 2.30$\,\pm\,$0.16. We include a small color correction of -0.017 for 24$\mu$m, and the 70$\mu$m relation includes the updated aperture correction of +0.15 which were not applied in the original paper \citep{fad06,app04,fra06}. Indeed, most of the sources do not have this radio excess, and are therefore dominated by star formation.

\subsubsection{AGN Contamination is Low}

Using the three methods above, as well as the fits to SEDs discussed in Section~\ref{sedfit}, we identified only 30 possible AGN. One source was identified from three of the diagnostics: the Radio FIR correlation, the IRAC colors, and the SED fit;  three sources were identified by 2 methods (Radio FIR+BPT:1, BPT+IRAC:1, IRAC+SED:1); but most sources were identified by only one method (Radio FIR: 7, BPT: 14, IRAC: 3, SED:2). Therefore, star formation is the most important contributor to galaxy activity in the Coma galaxies of our sample. This represents a conservatively large estimate of the contribution of AGN activity, as the bolometric luminosity in these systems is not necessarily dominated by the AGN. For example, as discussed in Section~\ref{4p2}, the best fit SEDs immediately identify dominant AGN - only 4 of 210 confirmed Mid-IR members. Furthermore, \citet{vei09} derive a diagnostic using the ratio of 25$\mu$m/60$\mu$m flux for ultraluminous IR galaxies and Palomar-Green QSOs (PGQSOs). Using this diagram, only eight of our possible AGN correspond to$\,>\,$50\% AGN contribution, and 22$\,<\,$50\% AGN contribution. 

\subsection{Obscured Star Formation} \label{4p2}

At this point, we have used spectra to identify cluster members, and are confident that there is little contamination from AGN. We now quantify the amount of star formation by deriving the specific star formation rates based on the total infrared luminosity and the galaxy stellar mass. The total infrared luminosity is produced from dust heated by hot young stars, and as such provides a measure of {\it obscured} star formation.

\subsubsection{Spectral Energy Distributions}\label{sedfit}

\begin{figure*}
\includegraphics[scale=0.8,angle=0]{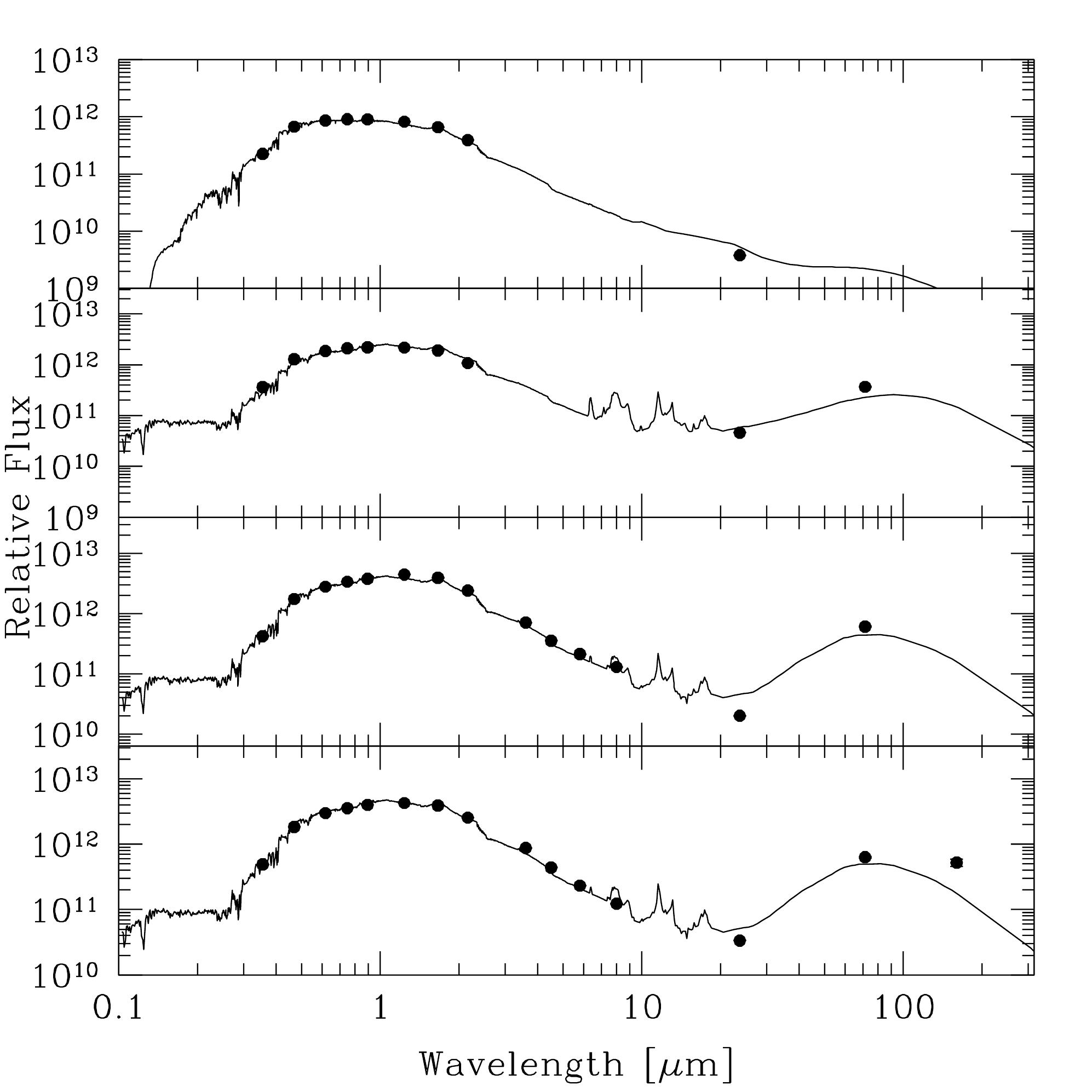}
    \caption[]{{\bf Sample best-fit template SEDs.} Measured photometric points are shown on top of the best-fit Polletta (2007) semi-empirical template SED.  }\label{sedpops}
\end{figure*}

We build the measured SED for the 210 confirmed 24$\mu$m emitting galaxies, using up to 15 photometric bands. These SEDs are compared to semi-empirical templates from \citet{pol07}, which are particularly well adapted to fitting infrared emitting galaxies. There are several templates for Elliptical, Spiral, Starburst galaxies, and AGN. These are based on results from the GRASIL \citep{silv98} code combined with photometric data from IR galaxies and AGN. The best-fit is the template with the lowest chi-square.  Figure~\ref{sedpops} shows a few examples of confirmed cluster members which have photometry in the representative bandwidths (optical, Near-IR and Mid-IR) and redshifts between 0.015-0.035. The total infrared luminosity is found by integrating the flux density between 8-1000$\mu$m of the best fit template. Only 4 of the 210 cluster members correspond to a best-fit template which is completely dominated by an AGN.

\subsubsection{Galaxy Stellar Mass}

To determine the galaxy stellar mass, we use the stellar model templates of \citet{mar05} with a Kroupa initial mass function \citep{kro01}, adopting a solar metallicity. These models describe the stellar emission from the UV to the Near-IR and we limit the fits to the optical and Near-IR portion of our observed SED, up to only IRAC 3.6 and 4.5$\mu$m. We modify the template SEDs in accordance with the \citet{cal00} extinction law, allowing E(B-V) to vary between 0 and 1. We then fit our observed SEDs to the host of available \citet{mar05} templates, and retrieve the stellar mass from the best-fit model. Photometric points from our Near-IR photometry, and those from the IRAC bands help to tightly constrain the SED fit around the stellar mass bump.

Table~\ref{elines} gives the calculated stellar mass for the 210 MIPS members.

\subsubsection{Specific Star Formation Rates}\label{ssfrsection4}

We convert the total infrared luminosity to a star formation rate (listed in Table~\ref{elines}) using the Kennicutt relationship \citep{ken98}. 

The specific star formation rate is simply the SFR divided by the stellar mass. It is a useful parameterization since Coma is host to a variety of galaxies. For example, a dwarf starburst galaxy may have a much smaller SFR than a large passive elliptical, yet a higher value of sSFR. In most of our discussion, we present the ratio f$_{sb}$, defined as the sSFR$\times$$\tau$, the timescale over which the galaxy is assumed to form stars at the currently observed rate. We take $\tau\,=\,$100$\,$Myr following \citet{fad06}.  Starburst galaxies here are those with f$_{sb} >$0.1.

\begin{figure*}
  \epsscale{2.0}
        \plotone{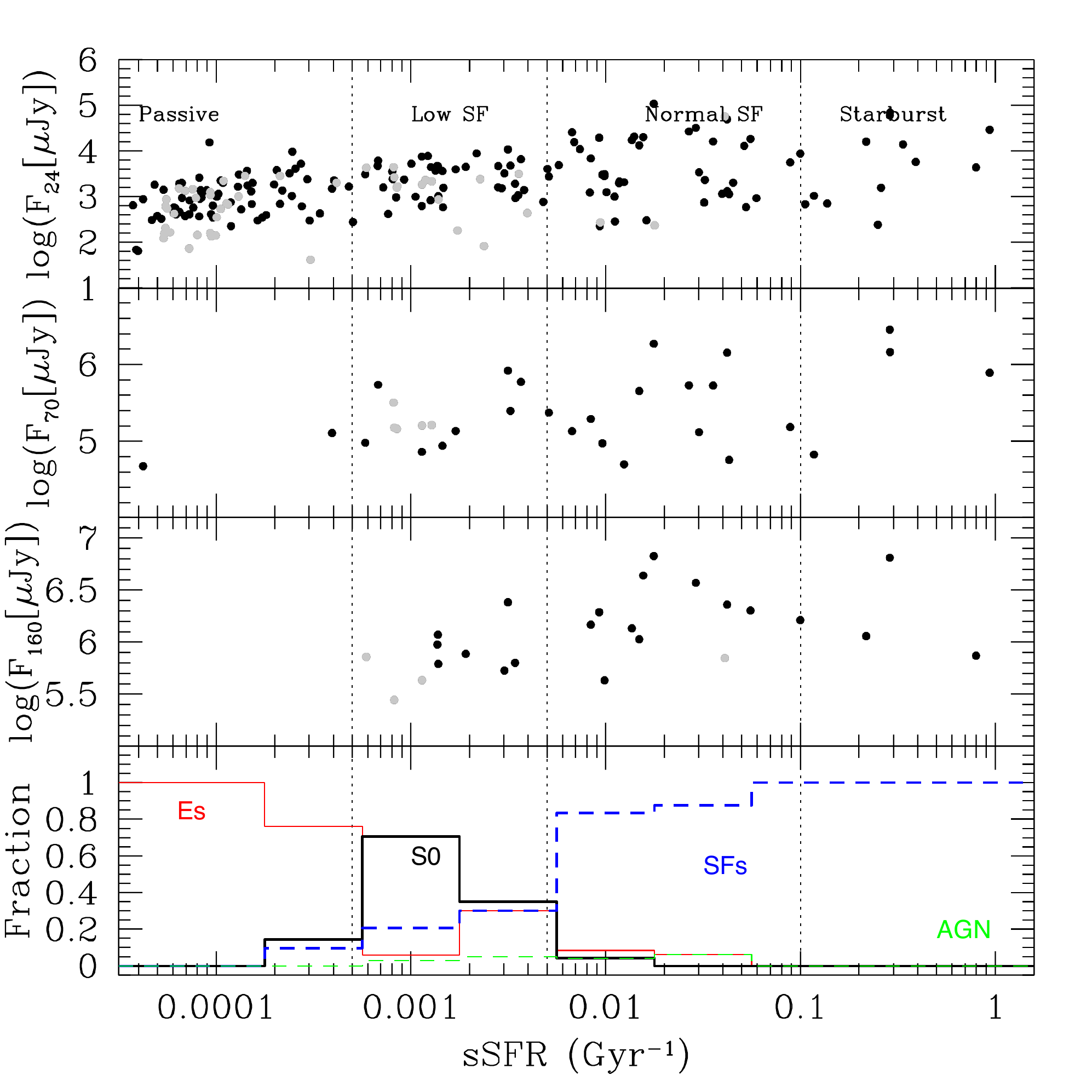}
    \caption{{\bf Specific star formation rates.} The Mid-IR luminosity of cluster members is plotted as a function of the specific star formation rate, derived from SED fitting. There are four regimes: Passive, Low star formation, Normal star formation, and Starburst. Grey points indicate member galaxies within 0.3$\,$degrees of the cluster X-ray centroid.\label{nspec} {\it Top} - The 24$\mu$m luminosity. {\it Second} - The 70$\mu$m luminosity. {\it Third} - The 160$\mu$m luminosity. {\it Bottom} - The fraction of galaxy types as a function of specific star formation rate. The galaxy types are those of the best fit templates from Polletta (2007). Galaxies are classified as ellipticals (thin red line), S0 (thick black line), spirals (thick blue dashed line), and AGN (thin green dashed line). Ellipticals dominate at low sSFR where red galaxies are the majority (see Figure~\ref{sfrcol}). Only four sources have best-fit templates corresponding to AGN.\label{sfrtype}}
\end{figure*}

Figure~\ref{nspec} shows the specific star formation rates with the galaxy flux at each of the Mid-IR wavelengths.  A range of 24$\mu$m luminosities give high specific star formation rates, suggesting that fitting the total SED, which finds both the mass, as well as the star formation rate, may provide a break in the degeneracy that exists when deriving star formation rates purely from the 24$\mu$m luminosity. 

We separate Figure~\ref{nspec} into four regions of sSFR. The passive galaxies have sSFR $<$ 5$\times$10$^{-4}$ Gyr$^{-1}$, low star formers have 5$\times$10$^{-4}$ $<$ sSFR $<$ 5$\times$10$^{-3}$ Gyr$^{-1}$, normal star forming galaxies have 5$\times$10$^{-3}$ $<$ sSFR $<$ 0.1 Gyr$^{-1}$, and starbursts have sSFR $>$ 0.1 Gyr$^{-1}$. Many of the MIPS-detected galaxies are passive, and the IR emission is likely a result of dust expelled by the large number of stars at later stages of their evolution \citep{tem08,cal10}.

The best-fit SED galaxy type, as given by the empirical SED, is plotted as a histogram with specific star formation in Figure~\ref{sfrtype}. The Elliptical, S0, and Spiral galaxy types correspond well to the expected behavior of the designations we set based on the sSFR, with passive ($<$ 5$\times$10$^{-4}$ Gyr$^{-1}$) corresponding to ellipticals. The few AGN are also displayed.

\section{Discussion}

We now turn to a discussion of the star formation rates based on SED fitting. In examining the galaxy colors, we find that the most red galaxies have the lowest sSFRs, and that a few galaxies with normal specific SFRs are also red.

We compare the sSFRs based on SED fitting, to values based purely on MIPS 24$\mu$m observations, confirming earlier results and extending to lower stellar mass. Additionally, we find that fits including 70$\mu$m data estimate slightly higher sSFRs than those from MIPS 24$\mu$m observations alone.

We compare the obscured sSFRs to extinction-corrected star formation rates derived from H$\alpha$ emission lines. As we will see, the extinction corrected unobscured values are able to completely recover the obscured values.

Finally, we describe the sSFR as a function of local galaxy density, finding a higher fraction of star forming galaxies at the cluster outskirts, than within the cluster core.

\subsection{The Color and Mass of Star Forming Galaxies}
 
\begin{figure*}
  \epsscale{2.0}
        \plotone{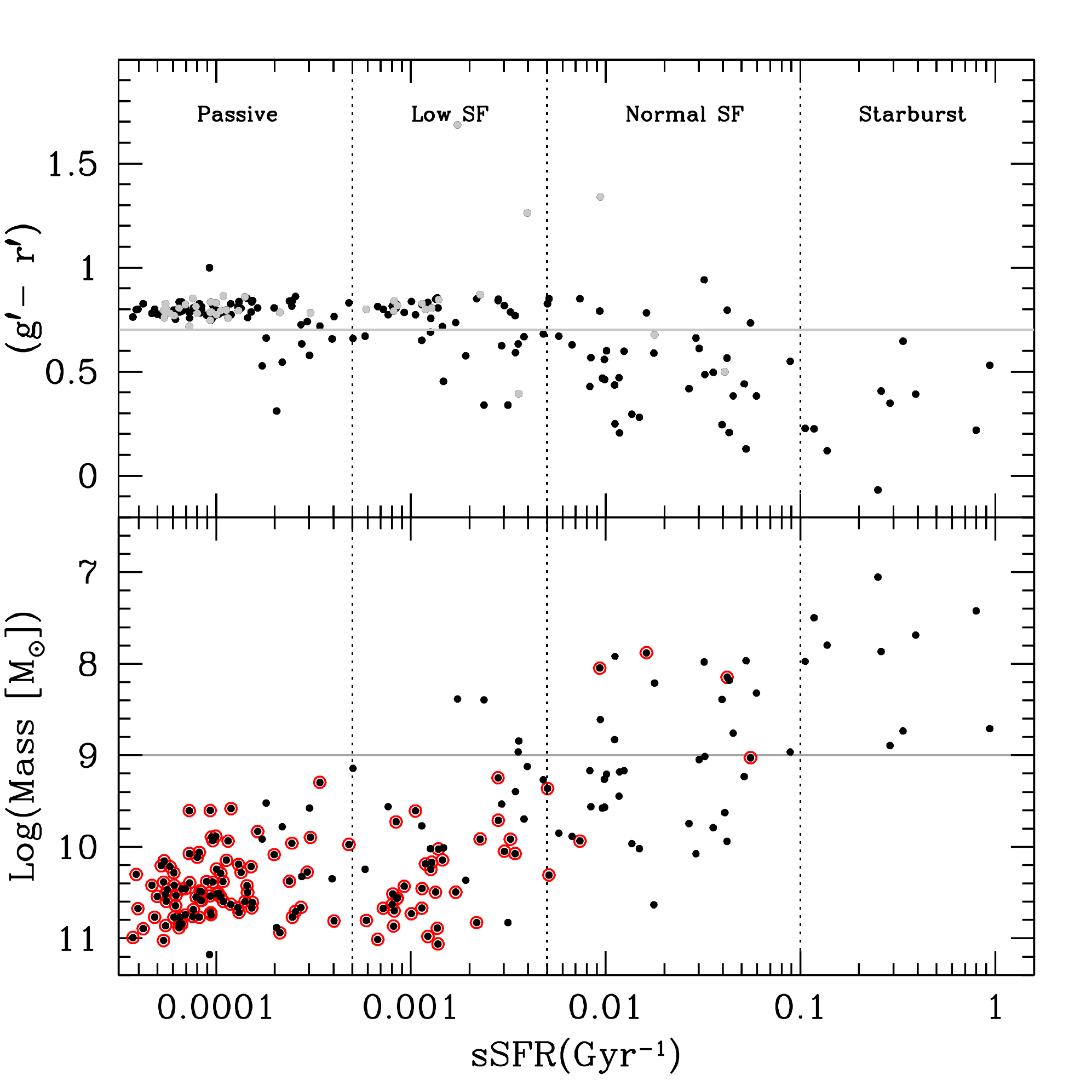}
    \caption{{\bf Galaxy Color and Mass.} The color and mass of Coma members are plotted as a function of the specific star formation rate as measured by SED fitting. There are four regimes: Passive, Low star formation, Normal star formation, and Starburst. {\it Top}  - The optical (g$^{\prime}$-r$^{\prime}$) color. The grey points highlight galaxies within 0.3$\,$degrees of the cluster core.  - {\it Bottom} The best-fit SED mass is shown, with a grey line showing our separation of dwarf galaxies with Mass$\,$$<$$\,$10$^{9}$M$_{\odot}$.  Red-sequence galaxies (see Figure~\ref{grcolmag}) are outlined in red. All of the starburst galaxies have blue colors with  (g$^{\prime}$-r$^{\prime}$)$<$ 0.7, and are dwarfs. \label{sfrmass} \label{sfrcol}}
\end{figure*}

Figure~\ref{sfrcol} shows the galaxy optical color and stellar mass, as a function of sSFR.  The starburst galaxies are blue, with (g$^{\prime}$-r$^{\prime}$)$<$ 0.7, and are dwarfs, with mass$\,$$<$$\,$10$^{9}$M$_{\odot}$. Most sources with a specific star formation rate less than 0.005 are red and have a large stellar mass. These are the red-sequence galaxies that dominate the cluster core and the FIR emission is likely caused by a large population of old red giant branch stars (also known to emit at MIPS 24$\mu$m; \citet{tem08,cal10}). 

\subsubsection{No Evidence for Dusty Starbursting Dwarfs}

One might expect that if a large population of dusty dwarfs existed, some low mass galaxies should be red by virtue of their dust extinction, rather than because of age. There are a few red dwarfs with normal star formation, but we do not detect any starbursting red dwarfs.

\begin{figure*}
  \epsscale{1.0}
        \plotone{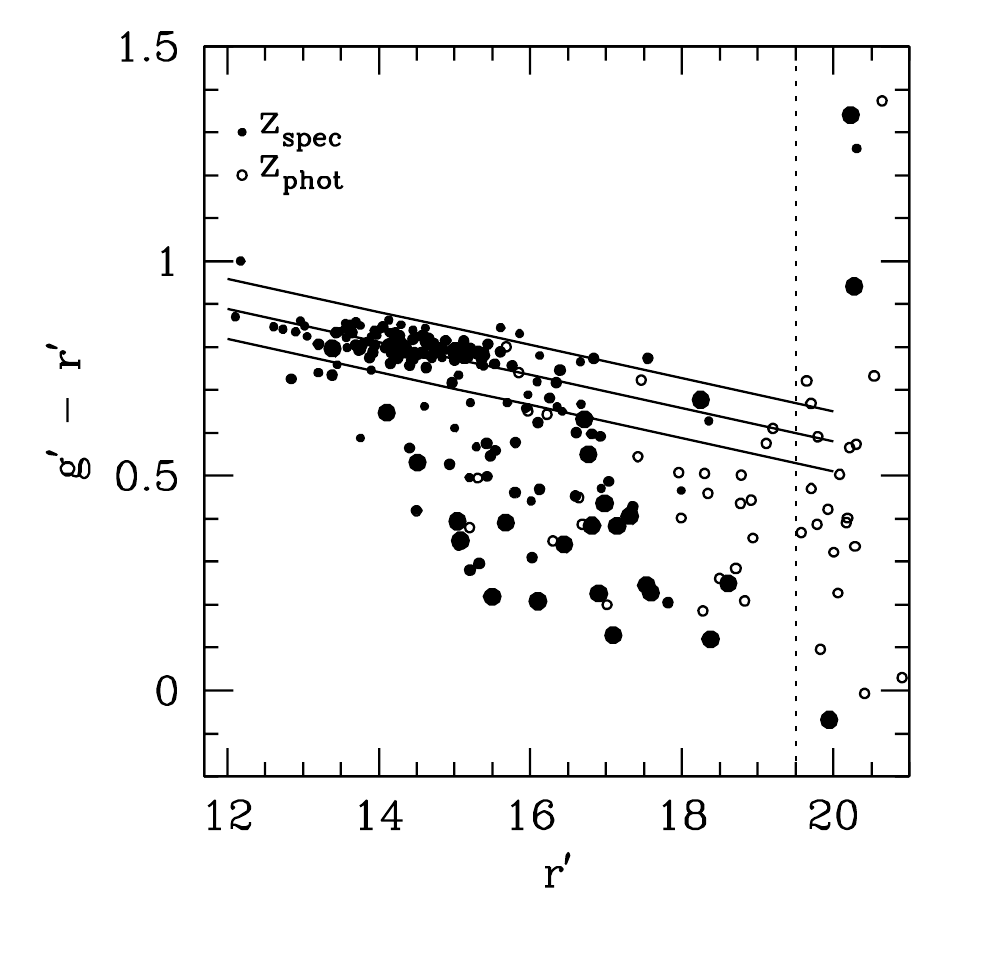}
    \caption{{\bf Optical color magnitude diagram.} The color magnitude diagram of Coma members with MIPS detections. Photometric members (with z$_{phot}$$<\,$0.08) are plotted as open circles and show the depth of the photometry. Spectroscopic members are filled black points. Large dots represent the dwarf galaxies. Our spectroscopic limit of r$^{\prime}$=19.5 is delineated with a vertical dotted line. There are no low mass red galaxies in the spectroscopic sample. There is also no evidence from the photometric sample that a significant population of red members exists just below the spectroscopic threshold. Therefore, the lack of dusty red dwarf galaxies is not because of our spectroscopic limit or incompleteness. \label{grcolmag}}
\end{figure*}

Because dust-obscured starburst galaxies should be optically fainter than their unobscured, blue, counterparts, it may be that our spectroscopic data is too shallow or incomplete to detect these fainter dwarfs. To investigate this issue, we present  Figure~\ref{grcolmag} which makes use of a deeper, complete catalog of sources based on photometric redshifts. The figure plots the color magnitude diagram of MIPS detected members. The spectroscopic members (filled black circles) are compared to the photometric members (open circles; defined as those with z$_{phot} <\,$0.08).  If it was true that our spectroscopic limit introduced a bias against dusty red star forming dwarfs, than the photometric members, which are much fainter, would show a sample of low-mass red galaxies. But, as the figure shows, there does not appear to be a missing population of red 24$\mu$m sources just beyond our spectroscopic limit of r$^{\prime}$$\sim$19.5. There is a caveat. Unfortunately, it is not possible to interpret the data below r$^{\prime}$$\sim$20.5 as the photometric redshifts become unreliable. Nevertheless, the lack of dusty red dwarfs does not appear to be caused by the spectroscopic limit of our study.

\subsection{Comparison of Star Formation Rates from other Methods}

\subsubsection{Obscured versus Unobscured Rates}

To derive the star formation rate for the unobscured component we measure the H$\alpha$ emission line and use the conversion of \citet{ken98}.

The H$\alpha$ emission was observed through fixed-diameter fiber spectrographs, but Coma galaxies are relatively nearby, so many of the galaxies are larger than the $\sim$3$^{\prime\prime}$ fiber diameter. Thus, we apply an aperture correction.  Using SExtractor on the SDSS r$^{\prime}$ image, we retrieve the isophotal area for each source. We assume that the star formation is spread throughout the galaxy area, thus the ratio of the galaxy area to fiber area gives the aperture correction. Corrections for the largest Coma members can be substantial, up to a factor of ten, however, the median correction is $\sim$2.

Star formation rates based on optical emission lines suffer greatly from extinction by surrounding dust, and so a second important correction we apply to the H$\alpha$ flux measurement is an extinction correction. In most sources, we can  measure the H$\beta$ emission line which we use to find the Balmer decrement and calculate the amount of extinction. We calculate a significant E(B-V) with a median of 0.55 and standard deviation of 0.42. Extinction and aperture corrected star formation rates from the optical emission lines are given in Table~\ref{elines}.

\subsubsubsection{SED SFRs Reproduce Extinction Corrected H$\alpha$ Values}

\begin{figure*}
   \center
  \epsscale{2.0}
     \plotone{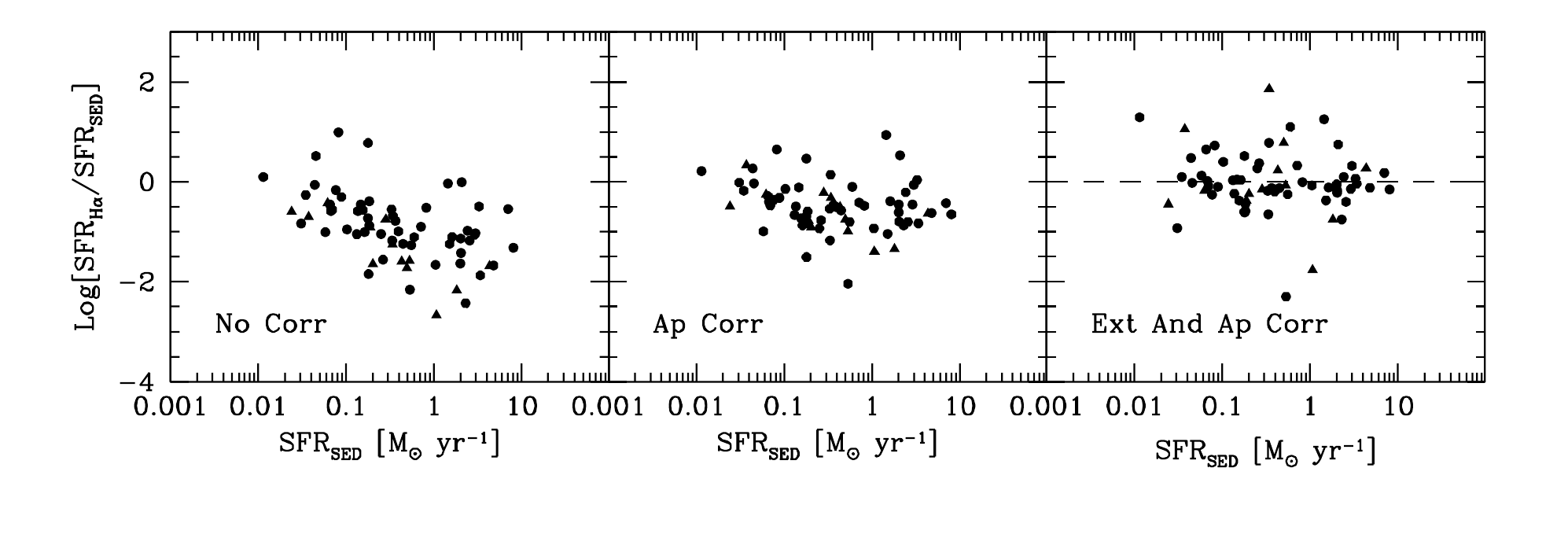}
    \caption{{\bf Unobscured versus obscured star formation rates.} The ratio of unobscured to obscured star formation rates as a function of obscured star formation rate. {\it Left} No correction is applied. {\it Center} - Only the aperture correction is applied. {\it Right} - Both aperture and extinction corrections are applied. Much of the optical emission in these nearby Coma galaxies is beyond the fiber diameter, so the aperture corrections are important (central panel). Extinction corrections raise the mean value of the sample. Once the extinction correction is applied, the H$\alpha$ based SFR matches the rate based on the total IR luminosity of the best-fit template. Any source that has been flagged as possible AGN in Section~\ref{agnsec} is shown as a triangle. There are few, and they inhabit no particular region of this diagram. 
\label{sfrouo}}
\end{figure*}

We can test how well the star formation rates based on total IR luminosities compare to values measured from corrected H$\alpha$ emission line fluxes. In Figure~\ref{sfrouo}, we show the ratio of the SFR derived from the optical emission line flux to that from SED fitting. 

 The aperture correction is more important for the larger galaxies, and once performed brings a uniform ratio of unobscured to obscured star formation for galaxies with significant SFR$_{SED}$ (Figure~\ref{sfrouo}, central panel). Including the extinction correction, performed on a source by source basis, the two rates approach each other. The median value of Log[SFR$_{H\alpha}$/SFR$_{SED}$] is 0.0, and the standard deviation is 0.5. This is an important point, for the sample of galaxies that have both emission lines and MIPS detections, the full star formation rates of the optical galaxies can be recovered after correcting for the extinction using optical emission lines. We suggest that this is because the same dust is causing extinction of the optical lines, as is producing the Mid-IR emission. 

\subsubsection{SFR Estimates from 24$\mu$m Flux only}

\begin{figure*}
\epsscale{2.0}
\plotone{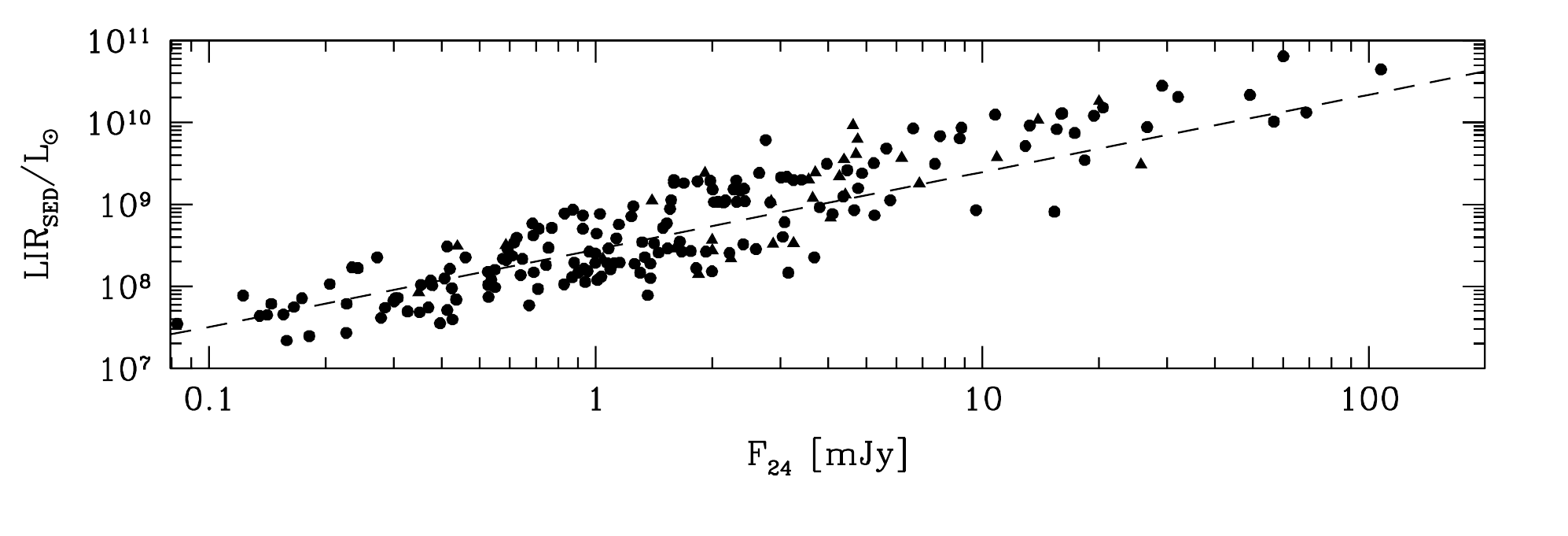}
\caption{{\bf Total infrared luminosity versus 24$\mu$m flux.} The total infrared luminosity from SED fitting is plotted against the measured 24$\mu$m flux. The dashed line shows the relationship developed by \citet{rie09}. Sources that have been flagged as possible AGN are shown as triangles. The overall agreement is good, however, beyond 4$\,$mJy the 24$\mu$m flux relationship underestimates the IR luminosity as measured by the best-fit SED.}\label{lirf24}
\end{figure*}

Many studies have used the 24$\mu$m emission alone as a proxy for total infrared luminosity \citep{rie09,bai06,mah10}. Certainly the SED fitting technique is more robust, as it is based on up to 15 photometric points. That the method is more accurate, is still open for debate, however, recent results from the {\it Herschel} space telescope support this. These new data show how star formation rates based on 24$\mu$m data underestimate rates derived by measuring longer wavelength fluxes, in 40\% of the cases \citep{raw10}. Our data also show a region where the 24$\mu$m data underestimates the rate calculated from the full SED fitting. In Figure~\ref{lirf24}, our results agree with the relationship of \citet{rie09} (the dashed line), out to the high flux end. Beyond 4$\,$mJy, the 24$\mu$m underestimates the IR luminosity as measured by the best-fit SED.

\subsubsection{Specific SFR from (24$\mu$m - z$^{\prime}$) Color}

In this section, we show that our results confirm and extend those of \citet{mah10}, who have recently explored the specific star formation in Coma cluster galaxies. These authors have used the (24$\mu$m - z$^{\prime}$) color as a proxy for quantifying sSFR. Our method relies on selecting galaxies based on the physical criteria of stellar mass and SFR. 

\begin{figure*}
\epsscale{2.3}
\plotone{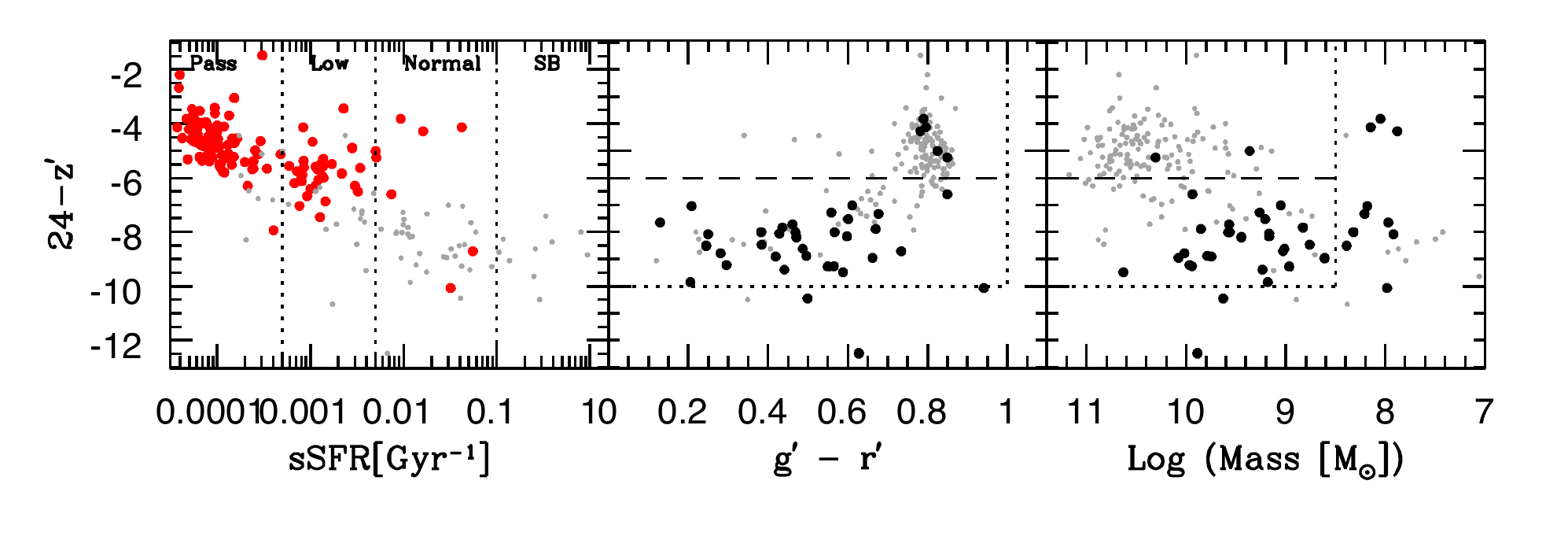}
\caption{{\bf The (24$\mu$m - z$^{\prime}$) color.} {\it Left} - We compare the specific star formation rate of cluster galaxies using two parameterizations.  On the y-axis we plot the (24$\mu$m - z$^{\prime}$) color, and on the x-axis, we plot the sSFR as determined from our SED-fitting technique. The vertical dotted lines define Passive, Low, Normal Star forming, and Starburst galaxies as in the text. The larger red points highlight red-sequence galaxies. {\it Center} - (24$\mu$m - z$^{\prime}$) color versus optical color, plotted to a (g$^{\prime}$ - r$^{\prime}$) of 1.0.  Here, and in the right panel, dotted lines show limits from \citet{mah10}, the horizontal dashed line is at (24$\mu$m - z$^{\prime}$) = -6, the separation used in \citet{mah10}, and larger black points indicate galaxies in the low-sSFR regime according to our SED fitting.  {\it Right} - (24$\mu$m - z$^{\prime}$) color versus galaxy stellar mass. We confirm the results of the \citet{mah10} study, and extend to lower stellar mass. This last is enabled by our deep Near-IR photometric points.}\label{m24z}
\end{figure*}

The first panel of Figure~\ref{m24z} shows that the usage of (24$\mu$m - z$^{\prime}$) recovers normal star forming and starbursts, galaxies with specific SFR $>$ 0.01 Gyr$^{-1}$, although, their color cut of -6 also includes contamination by many lower sSFR galaxies. There is also a population of galaxies with low and normal specific star formation rates (between 0.0005-0.1 Gyr$^{-1}$) which lie above the (24$\mu$m - z$^{\prime}$) color cut of -6. 

One of the results of our work is that the infrared star forming and starburst galaxies are blue, having (g$^{\prime}$ - r$^{\prime}$)  $<$ 0.7.  The central panel of Figure~\ref{m24z} confirms that the (24$\mu$m - z$^{\prime}$) $>$ -6 selects the same blue and red populations based on the optical color cut.

The right panel of Figure~\ref{m24z} shows that we are able to probe deep into the low-mass regime, measuring galaxies with stellar masses as low as 10$^{7.5}$M$_{\odot}$. The vertical dotted line shows the limit of the \citet{mah10} paper at a mass of 10$^{8.5}$M$_{\odot}$. As we saw in Figure~\ref{sfrmass}, the low mass dwarf galaxies are important as the highest specific sSFR galaxies, the starbursts, all have masses lower than 10$^{9}$M$_{\odot}$. 

\subsection{Environment}

Most of the imaging extends beyond the virial radius of Coma, parametrized by r$_{200}$, the characteristic radius for the cluster. We can therefore compare the galaxies in the cluster core, 0.2$\times$r$_{200}$, to those on the edge of the cluster center, beyond half of r$_{200}$. For Coma, r$_{200}$ has been measured to be 1.5h$^{-1}$Mpc \citep{gel99}. At the cluster redshift, this translates to r$_{200}$= 88$\,$$^{\prime}$. 

\subsubsection{Location}

\begin{figure*}
  \epsscale{2.2}
        \plotone{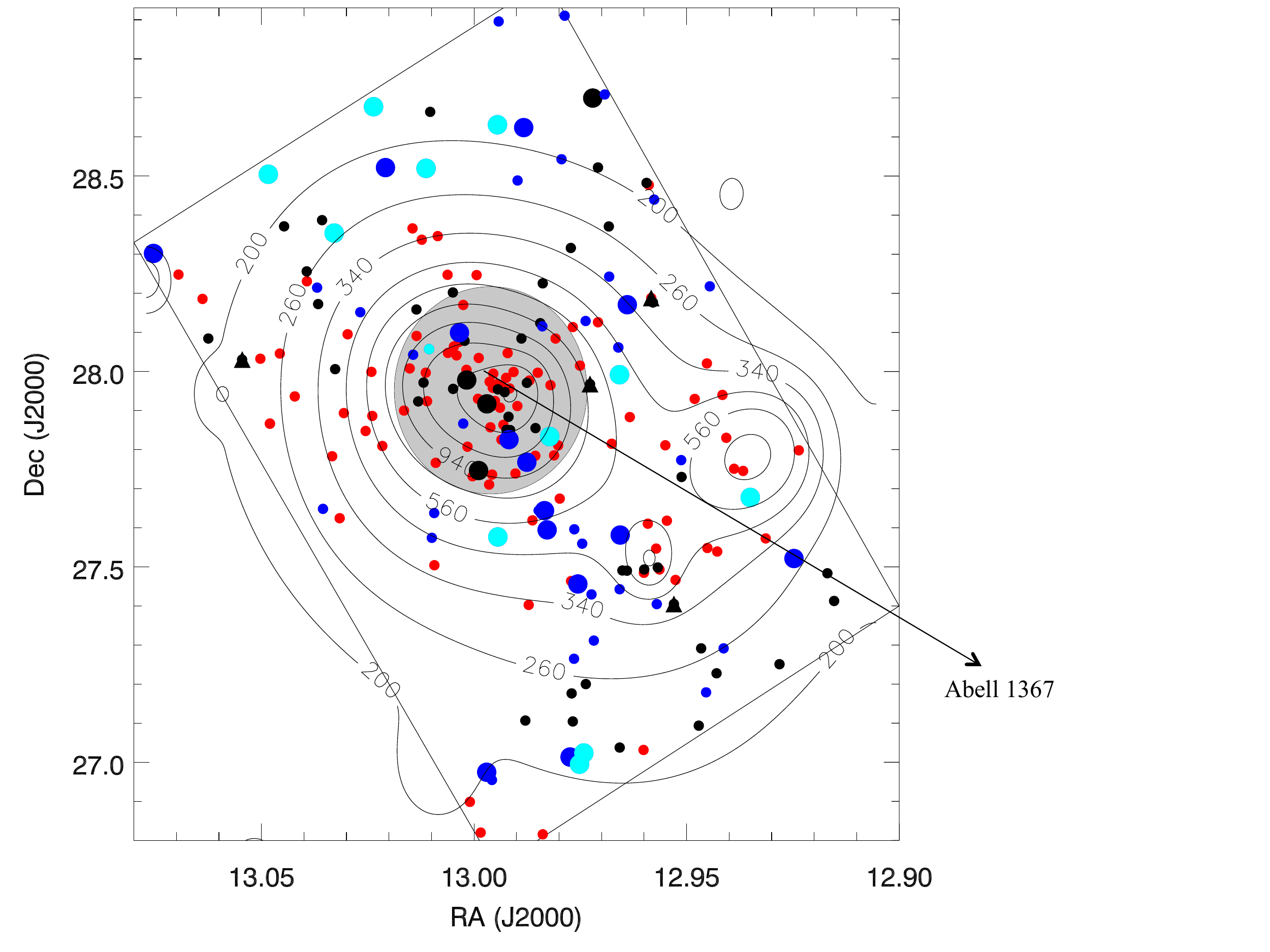}
    \caption{{\bf Spatial distribution of star formation.} The position of each spectroscopically confirmed MIPS-emitting members on the sky is shown. Red points are passive galaxies, black are low specific star formers, blue are normal star forming and cyan are starburst. The dwarf galaxies are shown as large points, and galaxies with masses greater than 10$^{9}$M$_{\odot}$ are smaller points. Seyfert galaxies are shown as black triangles. The grey circle has a radius of 0.3$\,$degrees and highlights the core region of the cluster, the black rectangle marks the MIPS field of view. Contours show the galaxy density, i.e., the number of red-sequence galaxies per square degree. As can be seen, the core has a dearth of starburst galaxies. \label{radec}}
\end{figure*}

Figure~\ref{radec} shows the positions of the spectroscopically confirmed 24$\mu$m emitting cluster members on the sky. Most dwarf galaxies are starbursts or normal star forming galaxies, and few exist in the cluster core. 

The suppression of star formation in the cluster core is also is consistent with \citet{cas01} who found a dearth of active galaxies in the core of the Coma cluster based on SDSS spectroscopy.

\subsubsection{Local Galaxy Density}

\begin{figure*}
  \epsscale{2.0}
        \plotone{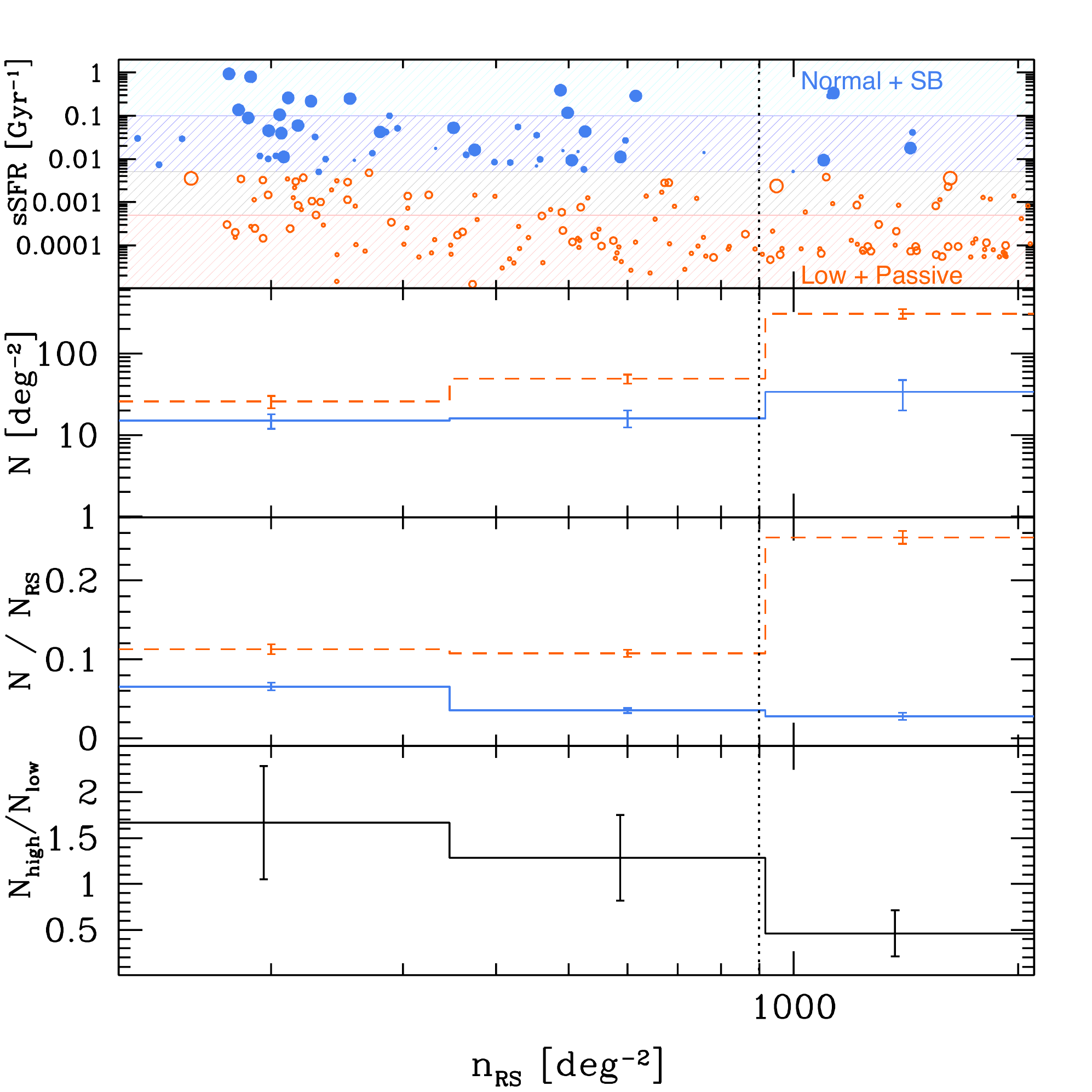}
    \caption{{\bf Local galaxy density.} {\it Top} - The specific star formation rate for cluster members as a function of galaxy density. The galaxy density is in units of number of red-sequence galaxies per square degree. A density of 900 corresponds to the core of Coma and is indicated with a dotted line. The filled blue circles represent galaxies with normal SF and as well as those currently undergoing a SB. The open red-orange circles represent galaxies with low and no star formation. The largest circle size showcases the Dwarf galaxy population, and the smallest circles are the massive red-sequence galaxies (as described in the text). For these last, we assume the MIPS emission is the old stellar population, and therefore, they are omitted from the histograms below.   {\it Second} - The number of galaxies per degree squared in bins of local galaxy density. The red-orange dashed line corresponds to low SF and passive galaxies, the blue solid line corresponds to normal and SB galaxies. {\it Third} - The number of galaxies normalized by the total density in the bin. There are fewer active galaxies throughout, with the largest fractional difference being in the core of the cluster, where passive galaxies dominate. {\it Bottom} -  The fraction of galaxies with high (sSFR$>$0.005 Gyr$^{-1}$) to low (sSFR$<$0.005 Gyr$^{-1}$) specific SFRs. The fraction of high-to-low star forming galaxies decreases toward the Coma core.   We show Poissonian errors on the histograms. \label{dens}}
\end{figure*}

We compute the galaxy density using the positions of 1557 red-sequence galaxies from the SDSS with 12$<$r$^{\prime}$$<$20, using a 2D wavelet transform as in \citet{fad98}. The red-sequence is the same as shown in Figure~\ref{grcolmag}, where (g$^{\prime}$-r$^{\prime}$) = 0.036$\times$r$^{\prime}$~-~1.320, and we include a range of $\pm$0.07 in (g$^{\prime}$-r$^{\prime}$) color.  The units of galaxy density are thus in number of red-sequence galaxies (N$_{RS}$) per square degree, represented as contours in Figure~\ref{radec}. The core and well-studied group NGC~4839 at (194.35, 27.5) are easily detected. A previously undetected group at (194.1,$\,$27.7deg) is also detected along the direction of the filament toward Abell~1367.

The top panel of Figure~\ref{dens} reveals a dearth of starburst galaxies in the cluster core, where the galaxy density n$_{RS}$$ < $300. In this figure, we aim to examine the star forming galaxies with both high, and low sSFRs. Thus, in the histograms we have excluded passive galaxies, those on the red sequence and having masses $>$10$^{10}$$\,$M$_{\odot}$. We also removed the 4 galaxies with SEDs best fit by AGN. In the second panel, the number of galaxies with a high sSFR (blue solid line) or a low sSFR (red-orange dashed line) is normalized by area. N increases toward the cluster core for both populations. The number of star forming galaxies with high specific SFR ($>$0.005 Gyr$^{-1}$) is lower than the number of low sSFR galaxies with the most differentiation in the cluster core. Even when normalized by the number of red sequence galaxies (third panel), the modestly star forming galaxies dominate over normal and SB galaxies, for all density bins.

Only a minority of passive galaxies are detected in the infrared, therefore, the number of red sequence galaxies (which may or may not have significant infrared emission), is much larger than the number of galaxies which we have defined as being passive from their low specific star formation rates.  The number of high specific SFR galaxies normalized by the number of low specific SFR galaxies is shown in the bottom panel. Indeed the trend is the same as for the solid blue line of N$_{SB}$/N$_{RS}$, with a progressive decrease toward the cluster core, but the numbers are quite different. We detect only $\sim$10\% of the red sequence galaxies as infrared, low star forming galaxies outside the cluster core, and roughly $\sim$30\% in the highest density regions of the core. This is consistent with a larger fraction of red sequence galaxies having been recently quenched in the cluster core. As a caveat, we reiterate that the observations of the Coma do not extend much past the virial radius, so we are not sampling the filament that connects to Abell~1367.

\subsubsubsection{The Distribution of Specific SFRs}

\begin{figure*}
  \epsscale{1.5}
        \plotone{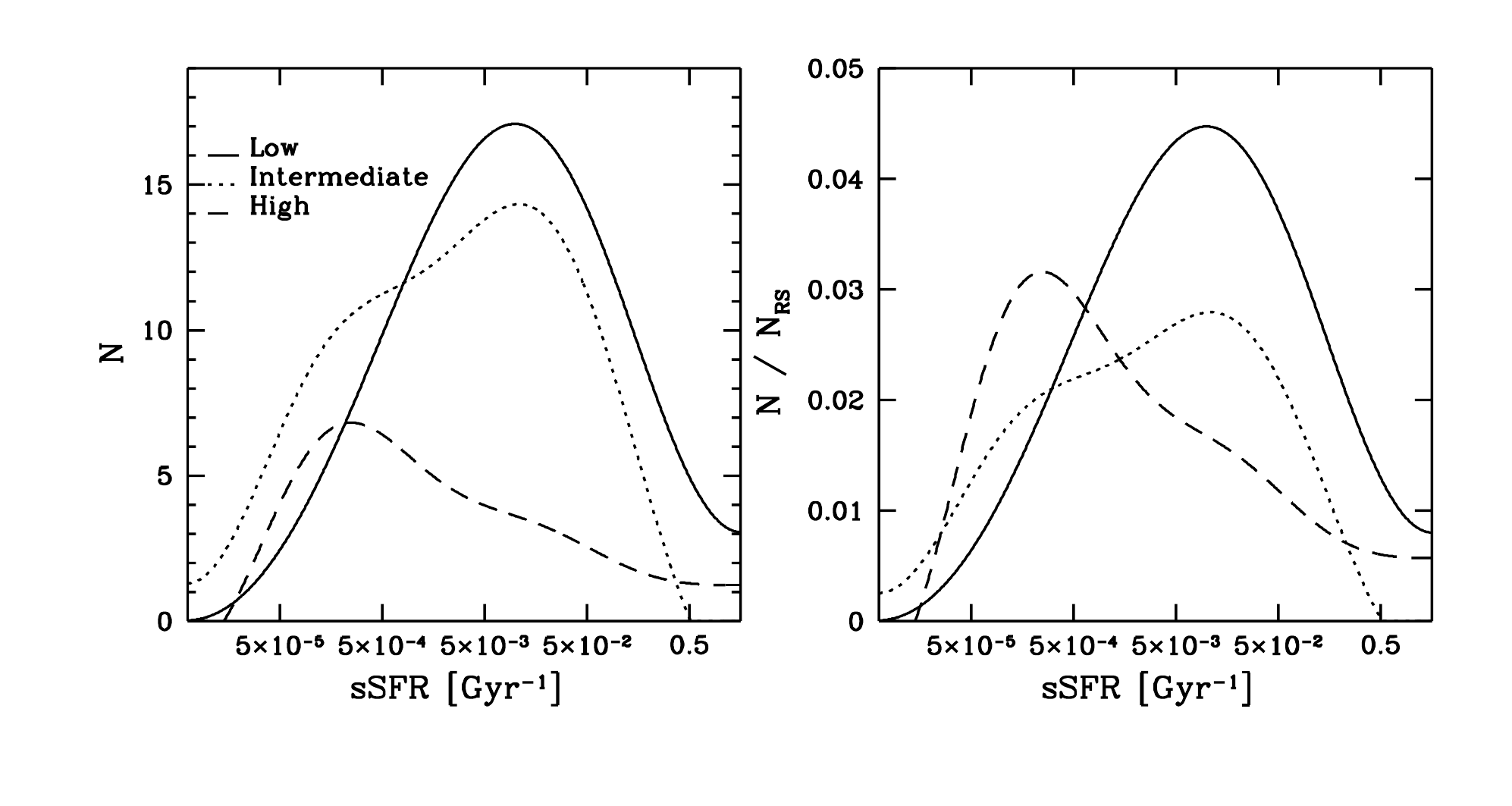}
    \caption{{\bf Distribution of sSFR.} The distribution of specific star formation rate is shown for the three density regions considered in Figure~\ref{dens}. The core, with density $>$900 is the dashed line, the intermediate regions are shown with a dotted line, and the lowest density regions with $<$300 are shown as the solid line. {\it Left} - The histogram of sSFR. {\it Right} - The distribution of sSFR for the three density regions, normalized by the total N$_{RS}$ in each bin. Large red galaxies, which likely have MIPS emission only from a population of old stars, are excluded from the distributions as are the 4 AGN identified from SED fitting. The galaxies in the high density core show a distribution of specific SFRs skewed to lower values. \label{ssfrdist}}
\end{figure*}

To further explore the effect of galaxy density, we examine the distribution of specific SFRs. The left hand side of Figure \ref{ssfrdist} shows the distribution of specific star formation rates for the core (dashed line; density $>$900), intermediate regions (dotted line; 300$<$ density $<$900), and the low-density outskirts (solid line; density $<$300). We again exclude AGN and large red galaxies, so the number of core galaxies is small. The right hand side of the figure shows the same, but normalized by the total density in each bin. The core galaxies have a distribution of sSFR skewed to lower values, peaking near 5$\times$10$^{-4}$ Gyr$^{-1}$, which, according to our SED fitting and Figure~\ref{sfrtype}, corresponds to where the fraction of elliptical galaxies begins to decrease. In other words, the core is dominated by a larger number of galaxies with low sSFR, as expected. The low density regions have a Gaussian distribution around 5$\times$10$^{-3}$ Gyr$^{-1}$ sSFR characteristic of quiescent (S0) and normal star forming galaxies. The intermediate density regions show both these populations, this may be a zone of transition between the two types.

\section{Summary and Conclusions}

We have exploited a plethora of archival as well as original photometric and spectroscopic observations in order to study the specific star formation of infrared member galaxies of the Coma cluster. Of particular importance for SED fitting is the new Near-IR photometry we collected at Palomar, as well as the inclusion of the longer wavelength MIPS observations. We matched the 24$\mu$m photometric catalog to spectroscopy gathered from the SDSS and to our own Hydra/WIYN observations to confirm 210 cluster members. We show that the contribution of AGN in the sample is low, and build spectral energy distributions which we fit to templates in order to derive the total infrared luminosity and galaxy stellar mass. 

We compare these estimates of obscured star formation to unobscured star formation rates derived from optical emission lines, finding that the two values agree once the optical spectra have been corrected for aperture size and extinction. This provides confidence to studies restricted to ground based observational methods only. For the filament and outskirt regions, where background contamination is high, the Mid-IR remains a good tool for initially identifying star forming galaxies. 

By further exploring the optical colors and specific star formation rates based on total infrared luminosities, we find that the starburst galaxies are blue dwarfs. This confirms the results of \citet{mah10} and extends their results to lower mass galaxies using a more robust probe of galaxy SFR and a more accurate measure of mass.

Additionally, we find that the star formation rates based on MIPS 24$\mu$m data alone represent well the values derived from fits to the total SEDs, except at high fluxes, where rates are slightly underestimated compared to those based on the SED fitting technique. We conclude that multi-wavelength photometry, in combination with fitting to template SEDs provides the most robust estimate of specific star formation rates. This is because the infrared hump is better constrained by values beyond the maximum. It is also a good galaxy identifier, as AGN are immediately spotted from their SEDs.

Confirming much of the recent work on Coma and nearby clusters in general, we also find that most of the star forming galaxies lie outside of the core. In particular, we have found that the ratio of high sSFR galaxies to low sSFR galaxies is lowest at the cluster core. We have also found that the distribution of the sSFR is skewed towards lower rates for the galaxies in high density regions, even after removing luminous red galaxies. This proves the core to be dominated by a larger number of low-SF galaxies. Furthermore, the intermediate density regions show a bimodal distribution in sSFR, with peaks at the same rates of both the high and low density regions.  We suggest therefore that these intermediate density regions are the transition zone between the low and high sSFR galaxies, in line with the popular notion that star formation is quenched in galaxies before they enter the cluster core.

Finally, we provide to the community fully reduced images in J, H, K$_{s}$, IRAC 3.6, 4.5, 5.8, and 8.0$\mu$m, and MIPS 24, 70, and 160$\mu$m through IRSA and 338 new spectra of 24$\mu$m sources through the NASA Extragalactic Database.

\acknowledgments

We thank F. Marleau for assisting with planning and performing the observations at Palomar, and also thank T. Jarret for sharing his WIRC reduction software which we used for the J, H, and K$_{s}$ fields. We also thank Andrea Biviano and Florence Durret for thoughtful readings and comments on the draft stages of this paper.

Support for this work was provided by NASA through an award issued by JPL/Caltech. This work is based in part on observations made with {\em Spitzer}, a space telescope operated by the Jet Propulsion Laboratory, California Institute of Technology, under a contract with NASA.  Funding for the SDSS and SDSS-II has been provided by the Alfred P. Sloan Foundation, then Participating Institutions, the National Science Foundation, the US Department of Energy, NASA, the Japanese Monbukagakusho, the Max Planck Society, and the Higher Education Funding Council of England. The SDSS is managed by the Astrophysical Research Consortium for the Participating Institutions (see list at http://www.sdss.org/collaboration/credits.html). This publication makes use of data from the Two Micron All Sky Survey, which is a joint project of the University of Massachusetts and the Infrared Processing and Analysis Center/California Institute of Technology, funded by NASA and the National Science Foundation. The NASA Extragalactic Database (NED) is operated by the Jet Propulsion Laboratory, California Institute of Technology, under contract with the National Aeronautics and Space Administration.

{\it Facilities:} Spitzer (MIPS), Spitzer (IRAC), Palomar 200in (WIRC), WIYN (Hydra).

\bibliography{a1763}

\end{document}